%% file: main.tex
\documentclass[12pt]{article}

\usepackage[round,comma,numbers,sort&compress]{natbib}
\usepackage{times}

\usepackage{url}
\usepackage{xcolor}
\usepackage{amsmath}
\usepackage{amsfonts}
\usepackage{hyperref}
\usepackage{booktabs}
\usepackage{graphicx}
\graphicspath{{j2v_figs/}}

\newcommand{\wv}{\texttt{word2vec}}
\newcommand{\tSNE}{\texttt{t-SNE}}

\newcommand{\si}[1]{Supplementary Materials #1}

\title{Neural Embeddings of Scholarly Periodicals Reveal Complex Disciplinary Organizations} 

\author
{Hao Peng,$^{1}$ Qing Ke,$^{2}$ Ceren Budak,$^{1}$ \\Daniel M. Romero,$^{1,3}$ \& Yong-Yeol Ahn$^{4,5,6*}$\\
\\
\normalsize{$^{1}$School of Information, University of Michigan}\\
\normalsize{$^{2}$Center for Complex Network Research, Northeastern University}\\
\normalsize{$^{3}$Center for the Study of Complex Systems, University of Michigan}\\
\normalsize{$^{4}$Center for Complex Networks and Systems Research,}\\
\normalsize{Luddy School of Informatics, Computing, and Engineering, Indiana University}\\
\normalsize{$^{5}$Indiana University Network Science Institute}\\
\normalsize{$^{6}$Connection Science, Massachusetts Institute of Technology}\\
\\
\normalsize{$^{*}$To whom correspondence should be addressed; E-mail: yyahn@iu.edu.}
}

\date{}

\begin{document} 

\baselineskip20pt
\maketitle 

\begin{abstract}
Understanding the structure of knowledge domains is one of the foundational challenges in science of science. 
Here, we propose a neural embedding technique that leverages the information contained in the citation network to obtain continuous vector representations of scientific periodicals. 
We demonstrate that our periodical embeddings encode nuanced relationships between periodicals as well as the complex disciplinary and interdisciplinary structure of science, allowing us to make cross-disciplinary analogies between periodicals. 
Furthermore, we show that the embeddings capture meaningful ``axes'' that encompass knowledge domains, such as an axis from ``soft'' to ``hard'' sciences or from ``social'' to ``biological'' sciences, which allow us to quantitatively ground periodicals on a given dimension. 
By offering novel quantification in science of science, our framework may in turn facilitate the study of how knowledge is created and organized.
\end{abstract}
\section*{Introduction}

Since the formalization of science, scholarly periodicals, such as academic journals and proceedings, have become the primary loci of scientific activities~\cite{garfield2006history, Fersht2009most, baldwin2015making, csiszar2018scientific}. 
Periodicals are not only the conduits of scientific communication, but also distributed repositories of scientific knowledge organized around topical niches and disciplines~\cite{merton1973sociology, csiszar2018scientific}. 
Therefore, scholarly periodicals have been considered the fundamental units for investigating the structure and evolution of science~\cite{boyack2005mapping, rosvall2008maps, bollen2009clickstream, borner2012design, uzzi-atypical-2013}. 

Moving beyond manually curated classification systems, previous studies leveraged citation and other metadata to capture relationships between periodicals in the form of similarity matrices or networks, which led to algorithmically-created ``maps of science'' and insights into the structure of disciplinary organizaition~\cite{small1999visualizing, boyack2005mapping, rosvall2008maps, bollen2009clickstream, borner2010atlas, borner2012design}.
Yet, although the vector-space model~\cite{salton1975vector} could provide a powerful framework for quantitative inquiries by allowing algebraic operations among periodicals, it has not been adopted much in the traditional approaches.  
The vector representation based solely on the explicit connections suffers from sparsity; for instance, using inter-citation or co-citation as a similarity measure produces a sparse matrix where most elements are zeros~\cite{boyack2005mapping}.
Incorporating indirect relationships would pose many choices for the metrics and require handling of a large, dense similarity matrix.

Recent advancement in machine learning has demonstrated that neural embedding techniques offer a powerful solution to these issues.
Neural embedding is a family of techniques for obtaining compact, dense, and continuous vector-space representations of entities that can efficiently encode multi-faceted relationships between those entities, and has become a core ingredient in modern machine learning~\cite{lecun2015deep}. 
The embedding approach, instead of focusing on the \emph{explicit relationships} between entities, aims to learn concise \emph{representations} that capture both explicit and implicit relationships between the entities.
Although its precursor, the vector-space model, was developed many decades ago~\cite{salton1975vector}, the neural network approach --- thanks to its flexibility, efficiency, and robustness~\cite{levy2015improving} --- has recently produced many breakthroughs. 

Since it was demonstrated that word embeddings can encode rich semantic relationships between words as geometrical ones in low-dimensional vector-space~\cite{mikolov2013efficient, mikolov2013distributed, mnih2013word, dong2017metapath, an2018semaxis}, the embedding models have offered novel opportunities and solutions to challenging problems, including language evolution~\cite{hamilton2016diachronic, rudolph2018dynamic}, gender and stereotypes~\cite{bolukbasi2016man, garg2018word}, culture and identities~\cite{caliskan2017semantics, kozlowski2018geometry}, and even the prediction of  material properties~\cite{tshitoyan2019unsupervised}. 
Furthermore, the idea of training vector-space embedding using neural networks is not limited to words --- it has been adopted to other entity types, including sentences, paragraphs, documents, images, and networks~\cite{le2014distributed, karras2019style, perozzi2014deepwalk, node2vec-kdd2016}.

Here, we propose a network embedding method to learn dense and compact vector-space representations of periodicals from the paper citation network.
We show that the periodical embeddings can effectively encode the complex organization of knowledge in science, which allows us to perform novel quantitative analyses such as making cross-disciplinary analogies between periodicals.
Namely, we show that our dense embedding approach can (i) better capture similarity between periodicals than alternatives, (ii) produce a high-resolution map of disciplinary organization that can provide insights into the existing classification systems, particularly regarding interdisciplinary research areas, (iii) allow us to make meaningful \emph{analogies} between periodicals, and (iv) identify robust spectra of periodicals along conceptual dimensions such as the soft--hard science axis and the social--biological science axis.

Our embedding method builds on the \texttt{DeepWalk} and \texttt{node2vec} model~\cite{perozzi2014deepwalk,node2vec-kdd2016}, which are a direct adaptation of the \wv{} model in the context of networks. In this framework, random walks on the network are considered as ``sentences.'' 
Instead of using the network of periodicals, our method leverages the richer and higher-order citation network of \emph{papers} to learn the representations of \emph{periodicals} (see~\nameref{methods}). 

Let us sketch the key idea. 
Imagine reading a paper from a field that you are unfamiliar with. 
To understand this paper, you may need to read another paper from the reference list; which in turn may prompt you to read another earlier paper, taking you to a rabbit hole of a \emph{citation trail}. 
We hypothesize that such citation trails, created from references between papers, capture natural \emph{sequences} in the citation network. 
Now, by looking at the periodicals where each of the papers in the citation trail was published, we can obtain a trail of \emph{periodicals}. 
Here, we consider each periodical as a ``word'' and each trail as a ``sentence''. 
If we apply the \wv{} model to these ``sentences,'' it lets us learn embeddings that encode the semantic relationships among periodicals.
Similar to the case of word embeddings, periodicals with similar context in the citation trails would have similar vector-space representations. 
Note that, instead of using random walks on the citation network of \emph{periodicals}, we leverage richer and higher-order trajectories from the lower-level \emph{paper} citations to enrich the output embeddings (cf. information gained from higher-order trajectories~\cite{rosvall2014memory}).

\section*{Results}

We applied our method to a citation network of 53 million papers and 402 million citation pairs built from the Microsoft Academic Graph (see~\nameref{methods}).
As a result, we obtained a 100-\emph{d} unit vector for each of the 20,835 periodicals. 
Our embeddings offer natural ways --- i.e.,~the cosine similarity between vectors --- to measure similarities between periodicals. 
For instance, the two closest periodicals to \emph{PNAS} are \emph{Nature} and \emph{Science}, and the two closest periodicals to \emph{American Sociological Review} are \emph{Social Forces} and \emph{American Journal of Sociology} (see \si{Fig.~\ref{fig:simi}} for the top list and other examples). 

\subsection*{Validating the embedding space}

We compare our dense periodical embeddings (``\textit{p2v}'') with two citation-based vector-space models. Specifically, we construct an adjacency matrix representing the citation counts between 24,020 periodicals. 
The first baseline is a citation vector model (``\textit{cv}''), for which we assign a 48,020-$d$ vector to each periodical by concatenating its in-degree vector and its out-degree vector (both are normalized to the unit length). 
In contrast to our method that compresses the information contained in the paper citation network into low-dimensional dense periodical embeddings, this citation vector model makes use of the citation network itself and each periodical is represented by its citation pattern with respect to every other periodical.
The second baseline is a periodical similarity matrix obtained by applying the Jaccard similarity measure to the periodical citation matrix (``\textit{jac}''), which is the best model reported in~\cite{boyack2005mapping}. In this similarity matrix (also a sparse matrix), an entry $m_{ij}$ represents the total number of citations between periodicals $i$ and $j$, normalized by their total number of citations to other periodicals. Each periodical is then represented as its row vector.

We evaluate our embeddings against the two vector-space models and other baseline methods in three tasks: (i) capturing the similarities between pairs of journals in the same discipline, (ii) comparing the ranking of similar journals to that perceived by experts, (iii) predicting the discipline category for journals. We focus on 12,780 journals for which we have the discipline information through the matching with UCSD catalog (\si{Table~\ref{tab:disc-match}}).

\begin{figure*}[ht!] 
\centering
\includegraphics[trim=0mm 0mm 0mm 0mm, width=0.5\linewidth]{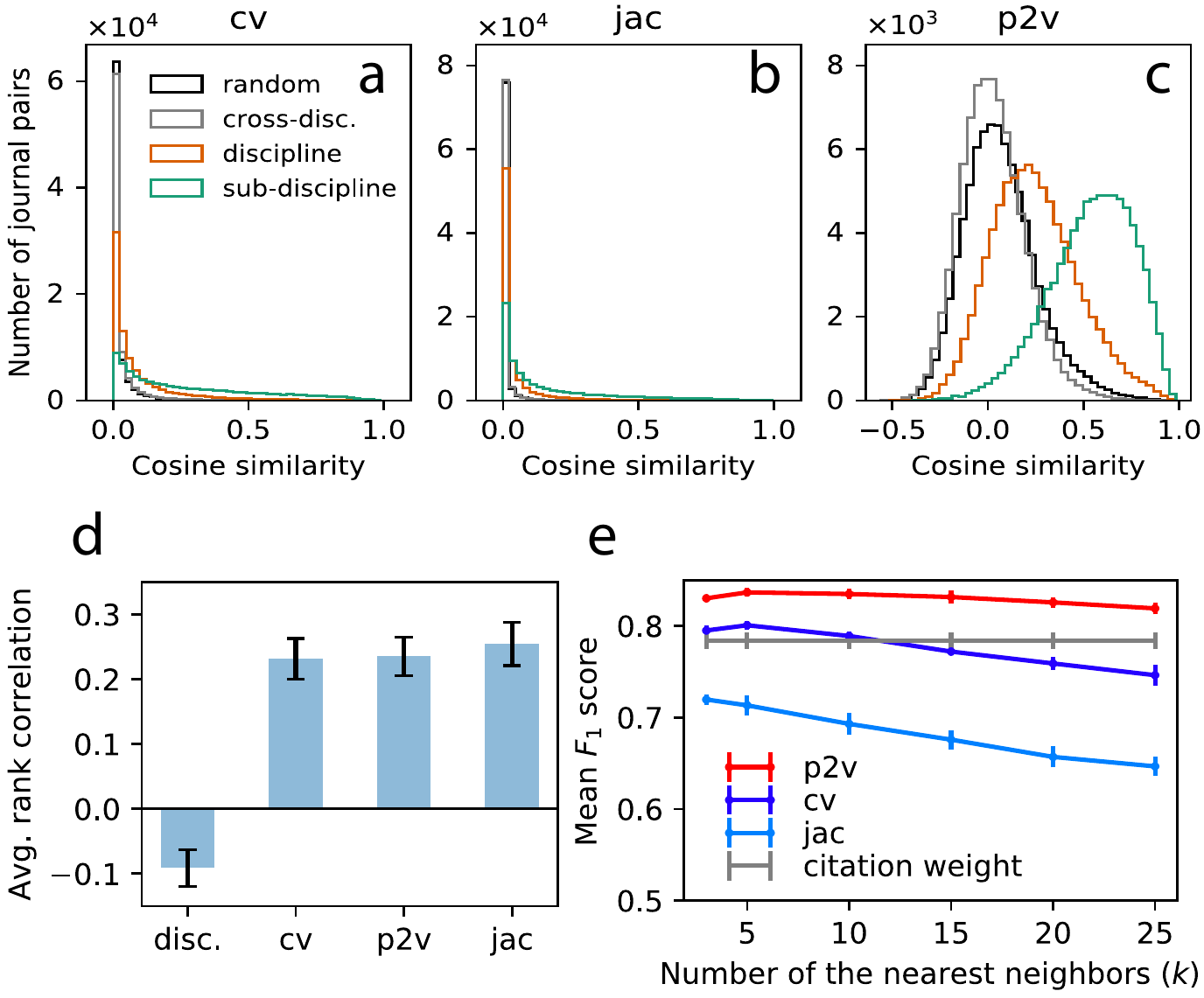}
\caption{\textbf{Model validation.}
\textbf{a-c}, The distribution of cosine similarities for four groups of 100,000 journal pairs calculated based on the citation vector model (\textit{cv}), the Jaccard similarity matrix (\textit{jac}), and our dense periodical embeddings (\textit{p2v}). The four labels --- \textit{random}, \textit{cross-disc.}, \textit{discipline}, and \textit{sub-discipline} --- represent random pairs, cross-discipline pairs, within-discipline pairs, and within-sub-discipline pairs. The two sparse embeddings (\textit{cv} and \textit{jac}) put most pairs at 0, and thus are not as informative as our dense embedding, which better captures journal similarities as well as their differences. 
Compared to random pairs, both the means and the distributions of the other three groups shift more dramatically based on \textit{p2v} than that based on either \textit{cv} or \textit{jac}.
\textbf{d}, The average rank correlation coefficient between algorithms and experts in ranking topically similar journals. Target journals with an average pairwise expert agreement above 0.2 are used in the evaluation. The label \textit{disc.} represents the method that ranks journals in the same discipline based on their PageRank scores.
\textbf{e}, The $F_1$ score of the classification task in predicting the discipline category for 12,751 journals (excluding 29 interdisciplinary journals) using the three vector-space models. The results are based on a 5-fold cross validation. The label \textit{citation weight} represents the method that predicts the discipline of a journal to be that of its most cited neighbor in the undirected journal citation network. Error bars indicate 95\% confidence intervals.
}
\label{fig:validation}
\end{figure*} 

The first task examines how well the embeddings can systematically capture journal similarities across disciplines. 
We randomly sample 100,000 journal pairs for four groups: (i) random pairs, (ii) pairs in different disciplines, (iii) pairs in the same discipline, and (iv) pairs in the same sub-discipline.
Figs.~\ref{fig:validation}a-c show the distribution of cosine similarities for journal pairs calculated based on the three vector-space models. 
According to the citation vector method and the Jaccard similarity matrix, most journal pairs in the same discipline (or even in the same sub-discipline) have a similarity score of 0 or close to 0 (the lowest possible value that can be produced by the two methods due to non-negative vector elements), highlighting the primary weakness of the sparse encoding approach: it fails to capture meaningful similarity variation across many pairs. By contrast, our embeddings provide a wide range of similarity scores (from -0.5 to 1) for random journal pairs (including pairs in different disciplines).

The mean similarity values for four groups of journal pairs are 0.02, 0.03, 0.10, 0.28 based on the citation vector model, are 0.008, 0.009, 0.046, 0.175 based on the Jaccard similarity matrix, and are 0.07, 0.03, 0.25, 0.54 based on our embedding (the corresponding mean values are statistically different from each other; $p$-values are negligibly small). 
We also compute Kullback-Leibler (KL) divergence by estimating the probability density function of similarity scores using the kernel density estimation (with the exponential kernel for \emph{cv} and \emph{jac}, and the Gaussian kernel for \emph{p2v}).
The distribution of within-discipline pairs shifts more dramatically from that of random pairs based on our embeddings (the KL divergence is 0.34 for \textit{p2v} vs. 0.25 for \textit{cv}; 0.11 for \textit{jac}). 
The displacement for within-sub-discipline pairs is even larger (KL divergence: 2.06 vs. 1.57 or 1.07), which is also true for cross-discipline pairs (KL divergence: 0.026 vs. 0.002 or 0.000).
This result underlines the benefits of dense and continuous embedding over sparse encoding in capturing journal similarities across disciplines defined in an existing journal classification system. 

Our second task is ranking periodicals based on their topical similarity to a given target periodical. In addition to our method (\emph{p2v}) and the two baseline vector-space models (\emph{cv} and \emph{jac}), we use another baseline (noted as ``\emph{disc.}'') that ranks periodicals in the same discipline based on their PageRank scores on the full directed and weighted periodical citation network. 
The rankings produced by these models are compared to a reference ranking that was constructed from an expert survey. 

The survey, which was distributed over the authors' institutions (see \nameref{methods}), asks experts to rank a set of candidate periodicals for a given target periodical based on their topical similarity. We then compare the algorithms' rankings with that given by experts. 
Fig.~\ref{fig:validation}c shows the average Kendall's rank correlation coefficient between each algorithm and experts. The three vector-space models perform similarly better than the first baseline. Although their performances are similar, our embeddings are orders of magnitude more computationally efficient than the citation vector model and the Jaccard similarity matrix in terms of time and space complexity due to low dimensionality (100 vs. 48,040 or 24,020).

The low correlation between algorithms and experts is mainly because experts themselves have high disagreement --- the average pairwise rank correlation per target journal is 0.14.
This may not be unexpected given the subjective nature of the task (see \si{Fig.~\ref{fig:survey-two}} for an example), which is also evidenced by the fact that on average 41.5\% of candidate journals were placed into the ``Unfamiliar Journals'' bucket by experts.

Finally, our third task tests the predictability of discipline category of a given periodical based on its neighbors. We focus on 12,751 journals (excluding 29 interdisciplinary ones).
We compare our embedding to the same citation vector model (\emph{cv}), the Jaccard similarity matrix (\emph{jac}), and another baseline (labeled as ``\textit{citation weight}''), which predicts the discipline of a target journal to be that of its highest-strength neighbor in the undirected journal citation network, where the edge weights are defined as the total number of citations between two journals (the undirected version performs better than the two directed versions). For \emph{p2v}, \emph{cv}, and \emph{jac}, we use the k-nearest neighbors algorithm based on vector similarities for the prediction task.
Fig.~\ref{fig:validation}d shows that our embeddings can more accurately predict journals' discipline category. In other words, the \emph{neighbors} in the embedding space tend to belong to the same discipline and this tendency is stronger in our model.  

Taken together, these results indicate that our periodical embeddings, while being much more efficient, can better capture the relationships between journals than the sparse vector-space models based on citations and other baseline approaches.

\subsection*{Disciplinary structure revealed by the periodical embedding}

\begin{figure*}[ht!] 
\centering
\includegraphics[trim=0mm 0mm 0mm 0mm, width=\linewidth]{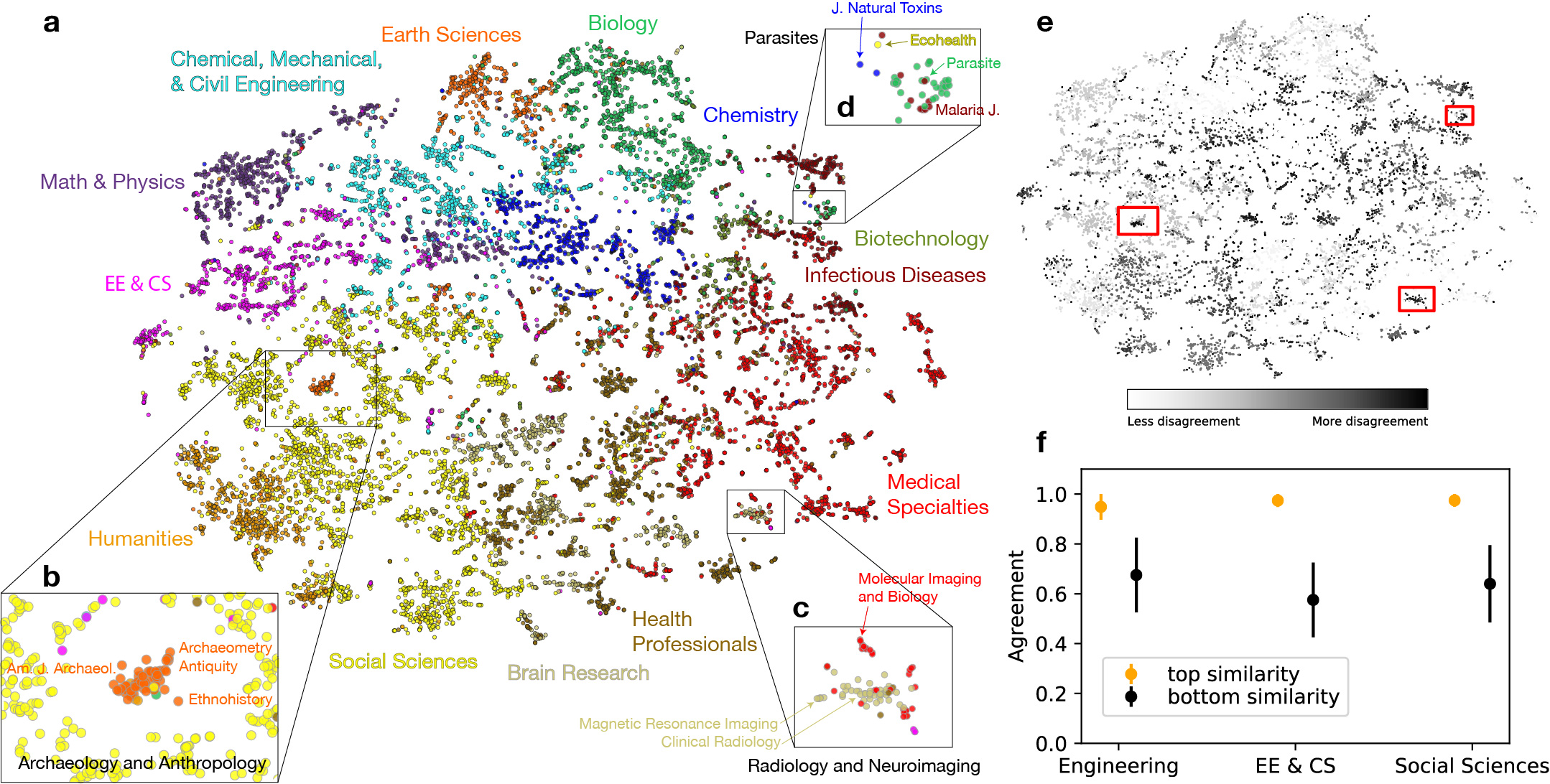}
\caption{\textbf{Periodical embeddings reveal complex disciplinary organizations.} \textbf{a}, The 2-\emph{d} projection of 12,780 journals obtained using \tSNE{}~\cite{maaten2008visualizing}. Each dot represents a journal and its color denotes its discipline designated in the UCSD map (29 multidisciplinary journals are colored in black).
\textbf{b}, archaeology and anthropology journals, classified as ``Earth Sciences'', form a distinct cluster with its center closer to ``Social Sciences'' than the major ``Earth Sciences'' cluster (verified by cosine distances). 
\textbf{c}, a group of medical imaging journals comes from ``Brain Research'', ``Medical Specialties'', and ``EE \& CS'', highlighting the key role of computer science and engineering in the study of brain imaging. 
\textbf{d}, a set of parasite-focused journals spans many disciplines, including ``Social Sciences'' (\emph{Ecohealth}), ``Biology'' (\emph{Parasites}), ``Infectious Diseases'' (\emph{Malaria Journal}), and ``Chemistry'' (\emph{Journal of Natural Toxins}), revealing the multi-faceted, highly interdisciplinary nature of parasite research.
\textbf{e}, The same map, but with a greyscale representing the level of disagreement between the clustering in our embedding space and the discipline categories in the UCSD map. Red rectangles highlight the locations in \textbf{b-d}.
\textbf{f}, The agreement between UCSD classifications and our survey. The top (bottom) represents journals with high (low) similarity between the UCSD catalog and a clustering based on our periodical embeddings.}
\label{fig:map}
\end{figure*} 

The embeddings of scholarly periodicals also encode the complex disciplinary structure in the knowledge space. Fig.~\ref{fig:map}a presents a 2-\emph{d} representation of the embeddings of 12,780 journals, providing an overview of the global structure of 13 major scientific disciplines (an interactive version is available at: \url{https://haoopeng.github.io/journals}).
Although our approach produces continuous --- not categorical --- representations of periodicals, to facilitate a comparison with a traditional journal classification system, we color each journal in Fig.~\ref{fig:map}a based on its discipline category designated in the UCSD map of science catalog~\cite{borner2012design}. 
The 13 disciplines defined in the UCSD map still show up as conspicuous regions in our projection. 
However, it also exposes the nuanced structure as well as the limitations of the classification approach. 
For instance, it uncovers interdisciplinary micro-clusters, such as parasite research or neuroimaging, that cannot be properly captured in the disjoint categories (see Figs.~\ref{fig:map}b-d and \si{Figs.~\ref{fig:map-social}-\ref{fig:map-mathphys}} for other examples).

If our embedding is indeed capable of capturing interdisciplinary periodicals, it is reasonable to hypothesize that \emph{stronger disagreement} about a periodical with a traditional classification indicates stronger \emph{interdisciplinarity} or \emph{wrong/ambiguous  classification}.

To test this hypothesis, we compare our vector-space map to the UCSD classification system (with 13 major categories) systematically and quantitatively. 
We apply the k-means algorithm to our embedding vectors and cluster journals into 13 groups (29 multidisciplinary journals were excluded from 12,780 matched journals). 
These organically discovered clusters are then compared to the 13 major categories in the UCSD classification system by employing the element-centric similarity measure~\cite{gates2019element}. 
This method allows us to quantify similarity between two clusterings at the level of individual element, thereby enabling us to quantify disagreement for each periodical.

Fig.~\ref{fig:map}e shows the map of agreement between the clustering based on our embeddings and the UCSD categorizations for 12,751 journals. Those interdisciplinary areas that we highlighted in Figs.~\ref{fig:map}b-d indeed exhibit strong disagreement.
The distribution of agreement scores for journals in each discipline is multimodal (\si{Fig.~\ref{fig:agree-hist}}). 
In other words, although the two clusterings are fairly similar for a large fraction of journals, there are still many whose membership across the two clusterings are distinct, possibly indicating their interdisciplinary nature. 
We performed a manual evaluation to estimate how clearly a periodical belongs to the discipline defined by the UCSD catalog for both high-agreement and low-agreement journals in three disciplines (see~\nameref{methods}). 


Fig.~\ref{fig:map}f shows that, journals with a high degree of agreement between the two clusterings can also be clearly identified in their designated discipline. On the contrary, for about 40\% journals on which the two clusterings strongly disagree, their discipline designation in the UCSD catalog is disputable. 
A manual inspection reveals that many low-agreement journals are interdisciplinary and difficult to be classified into a single category (e.g.,~\emph{Biostatistics}, \emph{Aggressive Behavior}, and \emph{Cell Biology Education} are classified as ``Social Sciences'' in the UCSD map).

These results suggest that our periodical embeddings, while agreeing with the UCSD categorization on clearly disciplinary journals (Fig.~\ref{fig:map}f, top), can identify interdisciplinary journals that are difficult to categorize into disjoint disciplines (Fig.~\ref{fig:map}f, bottom). 
This result shows that the dense periodical embedding is a promising data-driven approach to quantitatively operationalize interdisciplinarity using vector similarity. For instance, one may quantify a paper's degree of interdisciplinarity as the average cosine distance between its cited periodicals.



\subsection*{Cross-disciplinary analogies between scholarly periodicals} 

One of the primary reasons behind the wide adoption of the \wv{} model is its uncanny ability to capture semantic relationships geometrically in vector space~\cite{garg2018word, an2018semaxis, kozlowski2018geometry}. 
The most famous example goes like this: $\mathbf{v}(\emph{king}) - \mathbf{v}(\emph{man}) + \mathbf{v}(\emph{woman}) \approx \mathbf{v}(\emph{queen})$. 
That is, the difference between \emph{man} and \emph{woman} (or \emph{king} and \emph{queen}) vectors captures the axis of ``gender,'' which can be generalized to other gendered nouns such as \emph{brother} and \emph{sister} (i.e.,~$\mathbf{v}(\emph{brother}) - \mathbf{v}(\emph{man}) + \mathbf{v}(\emph{woman}) \approx \mathbf{v}(\emph{sister})$)~\cite{mikolov2013efficient, mikolov2013distributed}. 

Can we make similar analogies between scholarly periodicals using our embeddings? 
For instance, given a periodical pair $(A, B)$, where $A$ is a quintessential Computer Science periodical and $B$ is the one for Sociology, can $(\mathbf{v}(A) - \mathbf{v}(B))$ capture the axis that runs between Computer Science and Sociology?
If that is the case, given a ``seed'' periodical, we can also utilize the vector analogy to explore other periodicals that are closer to Computer Science and farther away from Sociology than the seed, or vice versa using the vector $(\mathbf{v}(B) - \mathbf{v}(A))$. 

\begin{figure*}[ht] 
\centering
\includegraphics[trim=0mm 0mm 0mm 0mm, width=0.75\linewidth]{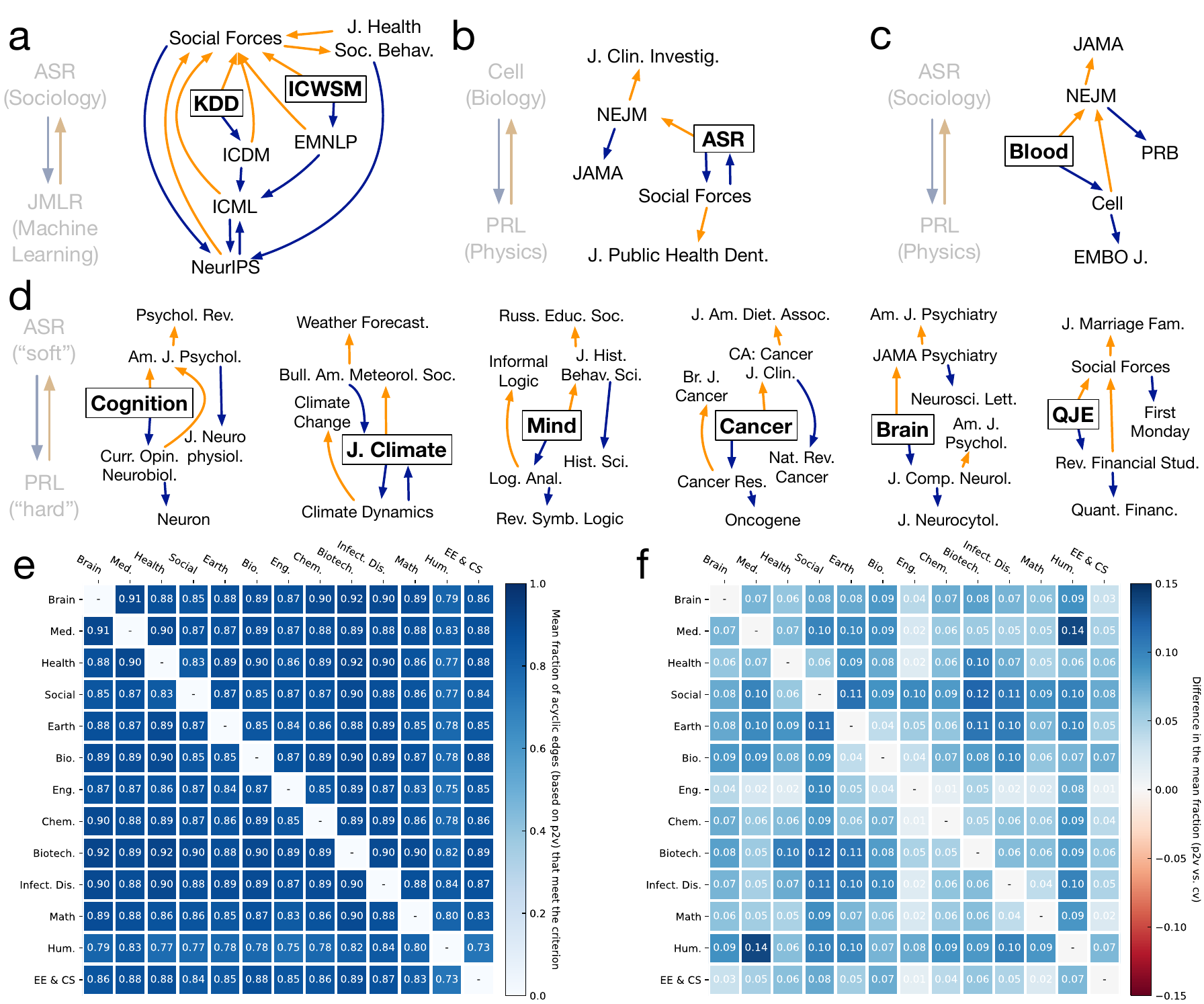}
\caption{\textbf{Analogy graphs between periodicals.} \textbf{a}, We apply two poles (\emph{ASR}, \emph{JMLR}) to \emph{KDD} (or \emph{ICWSM}) iteratively to find the most similar periodical at each step via the vector analogy: $\mathbf{v}(X) - \mathbf{v}(\emph{ASR}) + \mathbf{v}(\emph{JMLR}) \approx \mathbf{v}(?)$ (blue edges) or $\mathbf{v}(X) - \mathbf{v}(\emph{JMLR}) + \mathbf{v}(\emph{ASR}) \approx \mathbf{v}(?)$ (orange edges). 
Each node has two outgoing edges (blue or red) representing the two opposite analogies.
\textbf{b}, We apply (\emph{Cell}, \emph{PRL}) to \emph{ASR}, and only expand periodicals that are one step away from \emph{ASR} to make the graph concise. 
\textbf{c}, The graph obtained by applying (\emph{ASR}, \emph{PRL}) to \emph{Blood}. 
\textbf{d}, Similar to \textbf{c}, 
for seeds in different disciplines, including ``Brain Research'' (\emph{Cognition}, \emph{Brain}), ``Earth Sciences'' (\emph{Journal of Climate}), ``Humanities'' (\emph{Mind}), ``Medical Specialties'' (\emph{Cancer}), and ``Social Sciences'' (\emph{Quarterly Journal of Economics}).
\textbf{e}, The average fraction of acyclic edges per analogy graph that satisfy the author overlap criterion for all 1,800 analogy graphs (produced by our periodical embeddings, \textit{p2v}) in each of the 78 discipline pairs. \textbf{f}, Same as \textbf{e}, but for the differences in the mean values from the analogy graphs produced by \textit{cv}. For all discipline pairs, the difference is positive and statistically significant (at $p <$ 0.001).}

\label{fig:analogy}
\end{figure*} 

We would like to note that one needs to be cautious about the interpretation of word analogies. Commonly, word analogy does not allow duplicates (i.e.,~all words in the analogy need to be different), which can be misleading in some contexts such as the study of biases in word embeddings~\cite{nissim2020fair}. 
Here, we maintain this constraint because we specifically aim to discover a \emph{new} periodical using the analogy.

To demonstrate the possibility of making such cross-disciplinary periodical analogies, we create ``analogy graphs,'' which are constructed by repeatedly performing the vector analogy and taking the best candidate periodical at each step. 
We first choose two canonical disciplinary periodicals and consider them as the ``poles'' of an axis going from one discipline to the other. 
Using the two poles, given a seed periodical, we then iteratively make analogies to the seed and subsequently discovered periodicals. 
All identified periodicals, including the seed, can be visualized as a directed network with nodes representing periodicals and links representing the analogical relationships.

Fig.~\ref{fig:analogy}a shows the analogy graph for \emph{ICWSM} (\emph{The International AAAI Conference on Web and Social Media}) and \emph{KDD} (\emph{ACM SIGKDD Conference on Knowledge Discovery and Data Mining}), produced by applying \emph{JMLR} (\emph{Journal of Machine Learning Research}) and \emph{ASR} (\emph{American Sociological Review}) --- two poles of an axis that goes from Sociology to Machine Learning --- to each seed (\emph{ICWSM} or \emph{KDD}).
Fig.~\ref{fig:analogy}a reveals a spectrum of periodicals that sit between Sociology and Machine Learning, from a disciplinary sociology journal (\emph{Social Forces}) to interdisciplinary computational social science conferences (e.g., \emph{EMNLP} (\emph{Empirical Methods in Natural Language Processing}) and \emph{ICDM} (\emph{IEEE International Conference on Data Mining})), to more method-oriented machine learning conferences (e.g., \emph{ICML} (\emph{The International Conference on Machine Learning}) and \emph{NeurIPS} (\emph{The Conference on Neural Information Processing Systems})). 
Another analogy graph is obtained by applying the periodical pair (\emph{Cell}, \emph{PRL} (\emph{Physical Review Letters})) that represents the axis from Biology to Physics, to the seed journal \emph{ASR}, which identifies periodicals with biological flavor --- \emph{NEJM} (\emph{The New England Journal of Medicine}) --- or more physics flavor --- \emph{Social Forces} (Fig.~\ref{fig:analogy}b). 
We apply, in Fig.~\ref{fig:analogy}c-d, the pair (\emph{ASR}, \emph{PRL}) to periodicals across disciplines; for instance, when applied to \emph{Blood}, we can discover a more ``physical'' journal (\emph{Cell}) and a more ``sociological'' journal (\emph{NEJM}). Note that in Fig.~\ref{fig:analogy}d, we only identify the most similar periodical that is in the same discipline as the seed during each step.


We then more systematically examine the validity of the periodical analogies with an external dataset --- author overlap between periodicals. 
The intuition is that, as we move away from a periodical (say $A$) and towards another (say $B$) --- if the analogy works as intended --- we will arrive at a periodical that is farther away from $A$ but closer to $B$, in comparison with the original periodical that we started. 

Specifically, for a periodical analogy ``$A : B \sim C : D$'' (an edge ($C \rightarrow D$) in an analogy graph produced with \emph{A} and \emph{B} as two poles; Fig.~\ref{fig:analogy}), we verify if their author overlaps satisfy the following condition: $\frac{O(C, A)}{O(C, B)} > \frac{O(D, A)}{O(D, B)}$, where $O(P_1, P_2)$ is  the number of shared authors --- those who have published in both periodicals $P_1$ and $P_2$. 
That is, as one starts from \emph{C} and ends up at \emph{D} by moving further away from \emph{A} and getting closer to \emph{B}, we expect that the ratio of author overlap for $O(C, A)$ vs. $O(C, B)$ should be larger than that for $O(D, A)$ vs. $O(D, B)$. 

For example, for the analogy ``\emph{ASR} : \emph{JMLR} $\sim$ \emph{EMNLP} : \emph{ICML}'', the ratio of (\emph{EMNLP}, \emph{ASR}) author overlap to (\emph{EMNLP}, \emph{JMLR}) overlap should be larger than that between the (\emph{ICML}, \emph{ASR}) author overlap and the (\emph{ICML}, \emph{JMLR}) overlap.
We can then identify the acyclic edges that satisfy the criterion and obtain their fraction for any analogy graph (generated with a seed and two poles). 
Using this pipeline, we compare the quality of periodical analogies generated by our embeddings to that produced by the citation-based sparse encoding model defined previously.

We systematically generate, for each pair of disciplines ($D_1, D_2$), all 1,800 analogy graphs by selecting two poles and the seed from the top 10 journals in $D_1$ and $D_2$ (based on their PageRank scores; cf. ``\textit{disc.}'' in Fig.~\ref{fig:validation}c). For example, for the pair (``Social Sciences'', ``EE \& CS''), we have 10 journals in each field, which gives 100 pairs of poles; for each pair of poles, we have 18 journals ($10+10-2$) that can be used as the seeds; we can then generate 1,800 analogy graphs for (``Social Sciences'', ``EE \& CS'').
We calculate the average fraction of acyclic edges that satisfy the author overlap criterion for all 1,800 analogy graphs in each discipline pair (see \si{Fig.~\ref{fig:analogy-a}} for an example). 
We then compare the mean fraction for analogy graphs produced by the two vector-space models (\textit{p2v} vs. \textit{cv}; cf. Fig.~\ref{fig:validation}).
The results shown in Figs.~\ref{fig:analogy}c-d indicate that the periodical analogies produced by our embeddings are better aligned with author overlap between periodicals than those produced by the citation-based sparse vector model, for \emph{every} pair of the 78 possible discipline pairs. 

\subsection*{Extracting conceptual dimensions in disciplinary organizations} 

The power of embeddings to discover analogical relationships between periodicals prompts us to explore more general conceptual dimensions in the knowledge space, because the two disciplinary ``poles'' of a scientific ``axis'' can be defined not only by a periodical pair, but also by two sets of periodicals.
We first pick two general disciplinary areas and calculate their centroids by taking the average of all periodical vectors in each area. 
Given the two centroid vectors, we obtain an axis that runs from one disciplinary area to the other as we did in the previous examples with individual periodicals. 
Formally, let $\mathcal{S}^{+} = \{\mathbf{v}^{+}_1, \mathbf{v}^{+}_2, \dots, \mathbf{v}^{+}_m\}$ and $\mathcal{S}^{-} = \{\mathbf{v}^{-}_1, \mathbf{v}^{-}_2, \dots, \mathbf{v}^{-}_n\}$ be two sets of periodical vectors, the centroid of each set is computed as: $\overline{\mathbf{v}}^{+}=\frac{1}{m}\sum_{1}^{m}\mathbf{v}^{+}_i$ and $\overline{\mathbf{v}}^{-}=\frac{1}{n}\sum_{1}^{n}\mathbf{v}^{-}_j$. 
Then the axis vector is defined as: $\mathbf{v}_\text{axis} = \overline{\mathbf{v}}^{+} - \overline{\mathbf{v}}^{-}$. 
We measure the projection of a periodical $p$ to this axis using the cosine similarity between two vectors: $s(p, \mathbf{v}_\text{axis}) = \frac{\mathbf{v}(p) \cdot \mathbf{v}_\text{axis}}{\vert \mathbf{v}(p) \vert \cdot \vert \mathbf{v}_\text{axis} \vert}$. 
Here, we examine two spectra of scholarship: (i) ``soft'' to ``hard'' sciences~\cite{praise2005soft, cole1994sociology, hedges1987hard} and (ii) social sciences to life sciences.

\begin{figure*}[p] 
\centering
\includegraphics[trim=0mm 0mm 0mm 0mm, width=\linewidth]{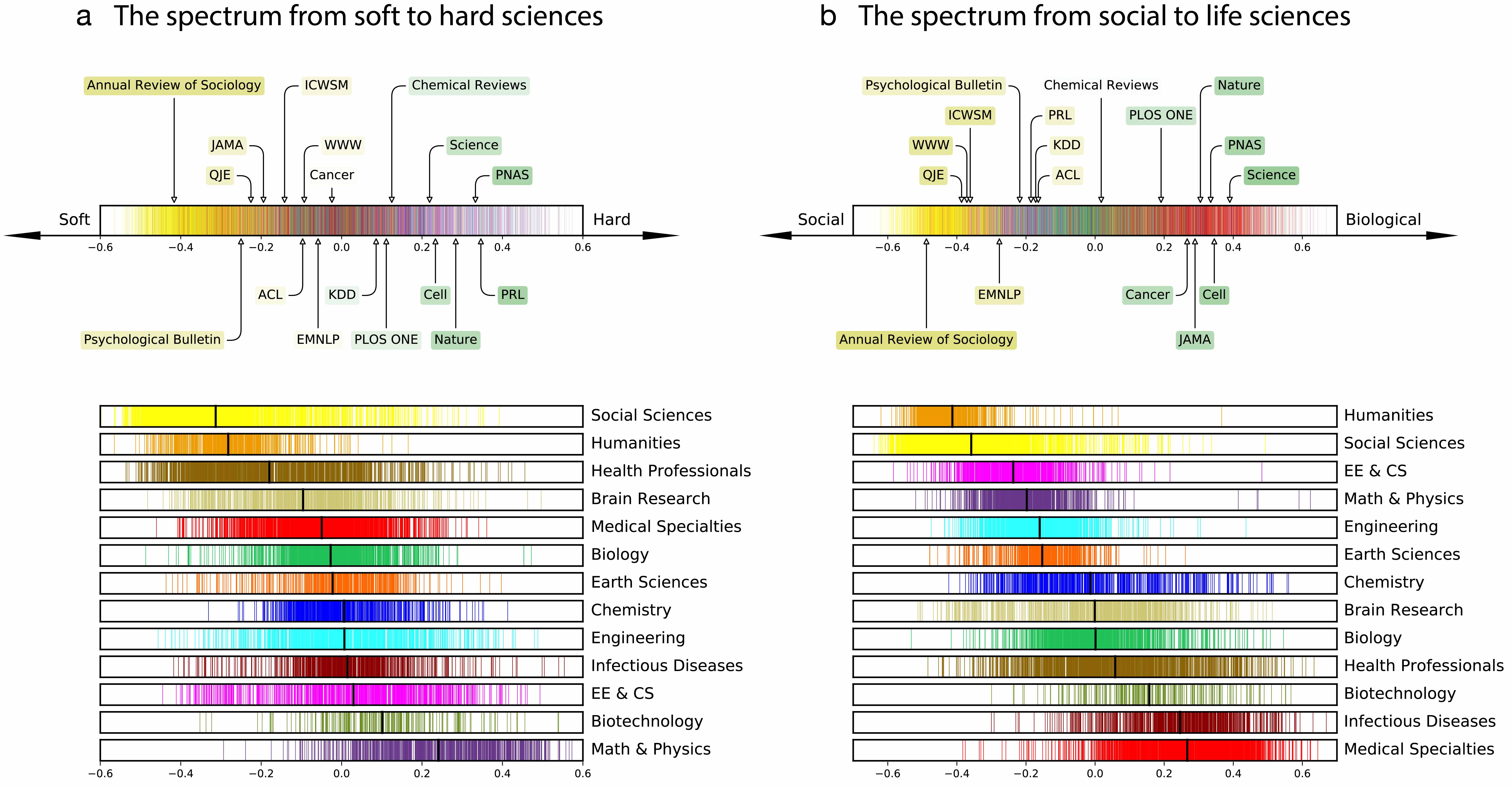}
\caption{\textbf{Two spectra of scholarship.} \textbf{a}, The spectrum of soft and hard sciences, operationalized by defining $\mathcal{S}^{+} = \{ \mathbf{v}(p)|p\in\text{``Math \& Physics''} \}$ and $\mathcal{S}^{-} = \{ \mathbf{v}(p)|p\in\text{``Social Sciences''} \lor p\in\text{``Humanities''} \}$. 
Each disciplinary journal is represented by a vertical line inside the box (12,751 in total). 
The color represents the discipline category and the position reflects the cosine similarity between the periodical vector and the axis $\mathbf{v}_{\text{soft}\rightarrow\text{hard}}$. 
We also annotate several journals and proceedings, whose background colors are proportional to their projection values. 
We then show journals in each disciplinary category separately at the bottom. 
The black vertical line in each discipline represents the mean projection value of its journals. 
\textbf{b}, The spectrum along the axis between social sciences and life sciences (biological), operationalized by defining $\mathcal{S}^{+} = \{ \mathbf{v}(p)|p\in\text{``Biology''} \lor p\in\text{``Biotechnology''} \lor p\in\text{``Infectious Diseases''} \lor p\in\text{``Health Professionals''} \lor p\in\text{``Medical Specialties''} \}$ and $\mathcal{S}^{-} = \{ \mathbf{v}(p)|p\in\text{``Social Sciences''} \lor p\in\text{``Humanities''} \}$.
Note that the ordering of 13 disciplines is dramatically changed from \textbf{a}, reflecting the complex organization of scholarly periodicals in the embedding space along scientific axes.}
\label{fig:spectrum}
\end{figure*} 

The first axis (dimension) captures the idea of the \emph{hierarchy of the sciences} --- an ordering of scientific disciplines by the complexity of the subject matter and the hypothesized order of development --- which places natural sciences like Mathematics and Physics at the bottom and social sciences like Sociology at the top~\cite{comte1855positive, cole1983hierarchy, fanelli2013bibliometric}. 
Disciplines at the top of the hierarchy are argued to be ``soft'' --- more complex, difficult to develop, and having less codified knowledge with more competing theories than disciplines at the bottom~\cite{cole1983hierarchy, hedges1987hard, lodahl1972structure}. 

We operationalize the axis from ``soft'' to ``hard'' sciences using two sets of periodicals. 
The pole of the ``hard'' sciences is defined by the centroid of all journals in ``Math \& Physics'' and the pole of ``soft'' sciences is defined by the centroid of all journals in ``Social Sciences'' and  ``Humanities'' (\si{Table~\ref{tab:disc-match}}). 
We project each periodical $p$ onto $\mathbf{v}_{\text{soft}\rightarrow\text{hard}}$ by calculating the cosine similarity $s(p, \mathbf{v}_{\text{soft}\rightarrow\text{hard}})$. 
The projection in Fig.~\ref{fig:spectrum}a forms a continuous spectrum along this axis, documenting how academic journals are distributed along the given axis that runs from Social Sciences \& Humanities to Mathematics \& Physics.

Some exemplary ``hard'' journals include \emph{Biophysical Journal}, \emph{Journal of Theoretical Biology}, \emph{Fractals}, \emph{Physics Reports}, and \emph{Physical Review E}. 
Some exemplary ``soft'' journals include \emph{Applied Psychology}, \emph{Anthropological Quarterly}, \emph{Law \& Society Review}, \emph{Sociological Forum}, and \emph{Politics \& Society}. 
Several representative periodicals are annotated in the spectrum. 
We also rank 13 disciplines by the mean projection value of all journals in each category in Fig.~\ref{fig:spectrum}a. 
The break-down into each discipline provides richer insights into how major scientific branches are organized along this conceptual dimension (see \si{Figs.~\ref{fig:soft-hard}-\ref{fig:spectrum-three}}). 
Overall, this spectrum shows that the ``hardness'' of academic disciplines increases in the order of Sociology, Psychology, Biology, Chemistry, Physics, and Mathematics, which concurs with the common conceptual ordering based on the hierarchy of the sciences~\cite{comte1855positive, cole1983hierarchy,xkcd}. 
\section*{Discussion} 

Here we present a continuous embedding framework for scholarly periodicals to systematically investigate the structure of periodicals and disciplines. 
By applying our method to a large bibliographic dataset, we obtain continuous and dense vector representations of scientific periodicals that can better encode the relationships between periodicals than two citation-based sparse vector-space models. The periodical embeddings can also offer new measurements that overcome conceptual and computational barriers. 
For instance, the framework allows us to make cross-disciplinary navigation using vector analogies, and to organize periodicals and disciplines along conceptual scientific dimensions.
More generally, the capacity to quantitatively operationalize relevant disciplinary dimensions will be useful for future studies that delve deeper into the complex disciplinary organization.

We acknowledge that there might exist comparable or better embedding methods given the rapid development in the field of machine learning~\cite{goyal2018graph}.
We view our primary contribution as one of the earliest attempts to fully embrace the embedding approach for the science of science by (i) developing a method that is motivated by the specific structure of scientific periodicals and papers, (ii) extensibly testing the usefulness of the embeddings with multiple evaluation tasks.

We also would like to point out limitations of our study.
First, the quality of embeddings depends on the quality of the dataset, thus our embeddings may reflect biases and errors in the data. 
For instance, it may be less useful for fields that are not covered by the source bibliometric dataset well. 
Second, the embedding approach suffers from sparse data; periodicals with fewer papers and citations will have embeddings that are less stable and less accurate, although it would still be better than only using explicit, direct links.
Third, as our method filters periodicals based on frequency (see~\nameref{methods}), it is possible that certain fields have fewer periodicals included in the embedding space. However, our matched 12,780 journals between MAG and UCSD catalog (\si{Table~\ref{tab:disc-match}}) indicate that every discipline still has at least several hundreds journals for the analyses, and the differences in coverage mainly result from their sizes according to the catalog (e.g.,~``Biotechnology'' has only 11 sub fields, but ``Social Sciences'' has 69 sub fields in the UCSD catalog).
Fourth, the embedding approach's assumption that vector representations are sufficient to capture explicit and implicit relationships between entities may not be valid. Indeed, it has been argued that such embeddings may be impossible to obtain for networks with high clustering~\cite{seshadhri2020impossibility}, although this may happen only in some embedding methods~\cite{chanpuriya2020node}. 
Similarly, when there exist multiple contexts for each entity~\cite{palla2005uncovering, ahn2010link}, a single vector may not be able to fully capture them.
In our case, the embeddings of multi-disciplinary journals may be skewed towards the primary disciplines that they publish and fail to capture explicit relationship to marginally published fields (see \si{Fig.~\ref{fig:simi}}a). 
Thus, when it is critical to consider explicit connections, the embedding approach may be inappropriate.
Furthermore, depending on the task, much simpler methods may well outperform embedding-based methods. 
For instance, a simple majority-voting approach outperforms our embedding in predicting the publication venue of a paper given its references (\si{Fig.~\ref{fig:pred-pub-venue}}).
Fifth, the present study does not take into account the evolution of periodicals and disciplines, falling short in providing a dynamic picture of the disciplinary patterns formed during different time periods. 
Sixth, although our method is more space- and time-efficient than the sparse vector models in downstream analyses, it does require non-negligible time and memory to train (e.g., a few hours to train the embeddings with 100 million citation trails) and thus applying the method to a larger dataset such as the universe of scientific papers can be challenging.
Finally, because the exact mechanisms and properties of neural embedding methods have not been fully understood, there may be biases that have not been understood. 

Despite these limitations, by demonstrating its validity and performance, we show that the embedding approach offers a promising avenue for science of science research. 
Future work may extend our framework to develop better embedding methods, investigate fundamental scientific questions with new capacities offered by the embeddings, or model the evolution of scientific periodicals and disciplines by incorporating temporal information in citations.

\section*{Materials and Methods} \label{methods}

\subsection*{Dataset} 


We used the Microsoft Academic Graph (MAG) data, which is the largest open access bibliometric dataset~\cite{sinha2015overview, alshebli2018preeminence}.
The snapshot we used contains 126,909,021 papers published in 23,404 journals and 1,283 conference proceedings between 1800 and 2016 (accessed on 02/05/2016).
There are 528,245,433 citations between these papers.

We focused on all papers that were published in either journals or conference proceedings, as the periodical information is needed to train the embedding model. 
Thus our study is based on a total number of 53,410,055 papers and 402,395,790 citations. 
They were published between 1800 and 2016 in 24,020 scholarly periodicals. 
\si{Fig.~\ref{fig:count}} shows the number of papers over time.

Using our method, we obtained embeddings for 20,835 periodicals (3,185 periodicals were dropped due to data filtering, see~\nameref{parameter}).
However, MAG does not have the discipline information for these periodicals. 
We thus used the UCSD map of science data, which contains discipline information for about 25,000 journals (classified into 13 academic disciplines). 
We matched 14,113 journals between MAG and the UCSD map based on journal names, among which 12,780 journals are covered in our embeddings (\si{Table~\ref{tab:disc-match}}).


\subsection*{Model} 

We consider the citation network between papers, where each node is a paper and a directed edge from $A$ to $B$ is formed if paper $A$ cites paper $B$. 
We generate many citation trails $ \{\mathcal{T}_1, \mathcal{T}_2, \dots, \mathcal{T}_N \}$ from the citation graph using random walks, where we first randomly choose a starting point (a paper) and randomly follow citations until we arrive at a dead-end (a paper without outgoing edges). 
Each trail $\mathcal{T}$ is a sequence of papers $(P^{\mathcal{T}}_1, P^{\mathcal{T}}_2, \dots, P^{\mathcal{T}}_{|\mathcal{T}|})$.
We discard trails that are immediately terminated ($|\mathcal{T}| = 1$). 
We then create a corresponding periodical trail $\mathcal{V}_{\mathcal{T}} =
(V^{\mathcal{T}}_1, V^{\mathcal{T}}_2, \dots, V^{\mathcal{T}}_{|\mathcal{T}|})$ for each paper citation trail, where the $i$-th element $V^\mathcal{T}_i$ is the publication venue (periodical) of the $i$-th paper $P^\mathcal{T}_i$ in the paper citation trail. 
Using the periodical trails, we learn two vector representations of each periodical $\mathbf{v}(V)$ (``input'') and $\mathbf{v}'(V)$ (``output'') by employing the skip-gram with negative sampling (SGNS) method~\cite{mikolov2013distributed}. 
For a given periodical citation trail $\mathcal{V}_\mathcal{T}$, the objective is to maximize the log probability
\begin{align}
O = \frac{1}{|\mathcal{V}_\mathcal{T}|}\sum_{t=1}^{|\mathcal{V}_\mathcal{T}|} 
\sum_{-w \le j \le w, ~j \neq 0} \log p(V^{\mathcal{T}}_{t+j} | V^{\mathcal{T}}_{t}), 
\label{obj}
\end{align}
where $w$ is the context window size. 
This training objective can be efficiently approximated as
\begin{align}
E = \log \sigma(\mathbf{v}'(V_O)^\top \mathbf{v}({V_I})) + \sum_{i=1}^{k}
\mathbb{E}_{V_i \sim \mathcal{U}(V)} \left[ \log \sigma(-\mathbf{v}'(V_i)^\top
\mathbf{v}(V_I)) \right],
\end{align}
where $V_I$ is the input periodical and $V_O$ is the output (context) periodical in Eq.~\ref{obj}, and $\sigma(x) = 1/(1 + \exp(-x))$. 
For each periodical pair $(V_I, V_O)$, SGNS samples $k$ negative pairs $(V_I, V_i)$ from the empirical distribution $\mathcal{U}(V)$. 
Here we let $k = 5$ and $\mathcal{U}(V)$ be the smoothed unigram distribution~\cite{mikolov2013distributed}. 
After training, the input vectors are used as the periodical embeddings~\cite{mikolov2013efficient}. 
All models are trained with $N = 100,000,000$. See~\nameref{parameter} for details.
SGNS method is efficient and scalable. 
It takes about 3 hours to train the embeddings with 100 million citation trails on a reasonably powerful computing server.
The algorithm for training embeddings is implemented in the \textit{gensim} package~\cite{rehurek_lrec}.

\subsection*{Hyperparameter tuning} \label{parameter}

We tuned two hyperparameters of the SGNS model: the context window size ($W$) and the number of dimensions ($D$). 
For each combination of $W$ (2, 5, and 10) and $D$ (50, 100, 200, and 300), we trained a model using the same 100 million periodical trails (\si{Fig.~\ref{fig:walk-length}}). 
We set the minimum periodical frequency to 50, which means that the embedding model will exclude periodicals with less than 50 occurrences due to data sparsity. 
A good model would output similar embedding vectors for periodicals that are similar in terms of research topics. 
We thus compared the quality of different embeddings based on the cosine similarities between periodicals.

Specifically, we randomly sampled 100,000 journal pairs for each of the three groups: (i) in the same discipline, (ii) in the same sub-discipline, and (iii) random pairs. 
Note that we focused on 12,780 journals for which we have discipline categories and are covered in our embedding model (\si{Table~\ref{tab:disc-match}}). 
\si{Table~\ref{tab:hyper}} indicates that the model trained with $W=10$ and $D=100$, which covers 20,835 periodicals, gives the best result. 
Fig.~\ref{fig:validation}c and \si{Figs.~\ref{fig:simi}}C-D show that, based on the best model, journal pairs in the same discipline (and sub-discipline) are much more similar in the embedding space than those selected randomly from any discipline.

\subsection*{Journal recommendation survey} \label{j-survey}

As an external evaluation, we use a survey to evaluate how well our embedding captures the similarity between periodicals.
We focus on the 12,780 journals that have discipline categories (\si{Table~\ref{tab:disc-match}}) because some baselines rely on disciplinary classification.
Each algorithm can rank, for a given target journal, the remaining 12,779 candidates. 
Note that the first baseline (\textit{disc.}) gives an arbitrary rank for journals whose disciplines are different from that of the target. 
We designed a survey to evaluate the three algorithms and recruited faculty members, researchers, and doctoral students from University of Michigan and Indiana University (The University of Michigan institutional review board guidelines were followed with human subjects). 
To make the task feasible, we selected top 20 journals in each discipline based on their PageRank scores. 
Journals belonging to the ``Interdiscipline'' category were excluded in the survey. 
For each of the 260 target journals, we constructed a set of candidate journals and asked participants to rank them based on their topical similarities to the target. The candidate set is the union of the top 4 similar journals given by each algorithm. Due to the overlap between the three top lists, the size of the candidate set varies between 4 and 12.

Participants first selected a discipline as their fields to begin the survey (\si{Fig.~\ref{fig:survey-one}a}). 
They were then asked about their familiarity with the 20 target journals in the selected discipline. 
Participants were allowed to continue the task only if they were familiar with at least three target journals (\si{Fig.~\ref{fig:survey-one}b}). 
Participants who selected less than three targets were immediately directed to the end of the survey. 
After the screening phase, participants were asked to rank, for each selected target, the set of candidate journals based on their topical similarities to the target. 
Participants can place unfamiliar candidates in the ``Unfamiliar Journals'' group (\si{Fig.~\ref{fig:survey-two}}).

Among 247 participants (out of 367) who finished the survey, 119 were qualified to complete the ranking task, and each of them was rewarded a \$10 Amazon gift card. \si{Table~\ref{tab:qualified}} shows the statistics of qualified responses across different disciplines.

Experts could give quite different ranking of the same target journal. 
We used Kendall's Rank Correlation coefficient $\tau$ to measure the level of agreement between two ranked lists of a target, based on the intersection of two ranked lists. 
We focused on target journals $\mathcal{J}$ whose average pairwise expert agreement $\hat\tau \geq$ 0.2. Note that this threshold is for determining which target journals should be used to evaluate the three algorithms.
In the evaluation step, in order to leverage more expert information, we appended to each ranked list the unfamiliar journals in a random order (the results are qualitatively the same without including unfamiliar journals).
Then, each ranked list for a target in $\mathcal{J}$ was used as the reference to evaluate three algorithms.
Specifically, for a ranked list $l_e^j$ of target journal $j$ from an expert $e$, we retrieved, from the full ranked list of an algorithm $a$, the order $l_a^j$ of journals in $l_e^j$, and we calculated $\tau_{(l_e^j, l_a^j)}$. 

The average correlation between each algorithm and domain experts (Fig.~\ref{fig:validation}d and \si{Fig.~\ref{fig:survey-j-three-disc}}) indicates that, across three disciplines, the three vector-space models are better than the first baseline, and are comparable to each other. 
The correlations between algorithms and experts are slightly higher with a higher threshold $\hat\tau$, but the error bars are also larger such that there is no clear winner between the three vector-space models (e.g., there are only 2 target journals with a total of 4 ranked lists for the evaluation with $\hat\tau$ = 0.8).

\subsection*{Evaluation with the UCSD categorization}

We test our hypothesis on the relationship between disagreement (with the UCSD catalog) and interdisciplinarity through a manual evaluation. 
First, we randomly selected 20 journals from the top 100 and the bottom 100 journals (10 from each) for three disciplines (EE \& CS, Engineering, and Social Sciences) based on the agreement score obtained by comparing the UCSD catalog with a clustering produced by our embeddings using the element-centric similarity measure~\cite{gates2019element}. We presented them to three of the authors and asked them to evaluate whether each journal belongs to the discipline defined in the UCSD categorization.
Each person was given the following instruction: ``Go to the journal's description page (via Google search or Wikipedia). Assign [yes] if only the target discipline is mentioned; Assign [no] if the target discipline is not mentioned; Assign [interdiscipline] if the target and other disciplines are mentioned; Assign [unsure] if no relevant information is found for this journal''. 

In the pre-test, the average pairwise agreement for 60 journals in three disciplines was 0.59 (Cohen's kappa), which is moderately high~\cite{mchugh2012interrater}. 
In the post-test, we again asked the three authors to evaluate another 20 journals for each of the three disciplines (each author evaluated a different set of journals).
We combined the pre-test responses (used the majority voting; 10 in the top and 10 in the bottom) and the post-test responses (30 in the top and 30 in the bottom) for each discipline. Journals with an [unsure] response were excluded in the analysis. We considered [yes] and [interdiscipline] as a journal being consistent with the UCSD catalog.


\subsection*{Validating the two spectra of science}\label{valid-axis}

We validate the spectrum of science in two ways.
First, to test the robustness of the two dimensions, we rebuild the axis vector by connecting the centroids of a subset of randomly selected journals in the two pole disciplines, and reorder all periodicals on the new axis. We then correlate this new ordering with their original arrangement (Fig.~\ref{fig:spectrum}). We repeat this process 100 times.
\si{Fig.~\ref{fig:axes-stability}} shows the average Spearman's rank correlation as a function of the number of journals used in the subset. The correlation is above 0.9 even when the new axis is built with less than 1\% of all journals in the field, for both the ``soft-hard'' axis and the ``social-bio'' axis. 
This high correlation suggests that the two axes are stable and robust.

Second, in our spectrum analysis, we used an axis anchored between the two centroids of two broad fields to score the fields themselves. We note that this could be problematic if the conceptual axis within the anchor field is not necessarily aligned with the overall axis, especially when the field does indeed exhibit such an axis internally, such as ``Social Sciences'' (\si{Figs.~\ref{fig:soft-hard}-\ref{fig:social-bio}}). 

To address this concern, we test whether the spectrum calculated at the level of the whole space is consistent with the spectrum calculated within the pole (anchor) discipline. 
Because it is unclear which sub-disciplines of ``Math \& Physics'' are ``softer'' or ``harder'' and the same issue seems to be true for Life Sciences with respect to the ``social-bio'' axis (\si{Figs.~\ref{fig:soft-hard}-\ref{fig:social-bio}}), we focus our efforts on social sciences. 

To validate the ``soft-hard'' axis in [``Social Sciences'' \& ``Humanities''], we first use ``Sociology'' as the soft subfield and ``Finance'' --- which has close connections to mathematics and physics --- as the hard subfield.
We rebuild the axis vector by connecting the two centroids and reordered all 20,835 periodicals on this new axis. 
Although we use only two sub-disciplines within a \emph{single discipline} to obtain scores for \emph{all} periodicals across \emph{all} disciplines, the Spearman rank correlation between the new ordering and the original one (Fig.~\ref{fig:spectrum}) is 0.73.
This result is also robust. 
We construct 9 soft-hard subfield pairs between three soft sub-disciplines (``Law'', ``Social Psychology'', ``Leadership \& Organizational Behavior'') and three hard sub-disciplines (``Finance'', ``Statistics'', ``Operations Research''). We then calculate the rank correlation between the ordering based on each pair and the overall ranking. The average correlation is 0.73 (95\% confidence interval: [0.69, 0.77]).

We repeat this robustness test for the ``social-bio'' axis in [``Social Sciences'' \& ``Humanities''], we use ``Sociology'' as the `social' subfield and ``BioStatistics'' as the `bio' subfield. 
The Spearman rank correlations between the new ordering and the original one is 0.82.
Similarly, the result is robust with other choices of subfields. 
We construct 9 social-bio subfield pairs between three social sub-disciplines (``Law'', ``Sociology'', ``Economics'') and three biological sub-disciplines (``Pyschiatric \& Behavioral Genetics'', ``Psychosomatic Medicine'', ``BioStatistics''). We then calculate the rank correlation between the ordering based on each pair and the overall ranking. The average correlation is 0.71 (95\% confidence interval: [0.63, 0.78]).

Altogether, these results demonstrate that the conceptual axis (either ``soft-hard'' or ``social-bio'') within the anchor fields is well aligned with the spectrum calculated with two broad disciplines, providing evidence that the two spectra are robust and meaningful. 



\bibliographystyle{unsrtnat}
\bibliography{main}

\clearpage
\noindent \textbf{Acknowledgements:} The authors thank Christopher Quarles, Xiaoran Yan, Cassidy R. Sugimoto, and Vincent Larivi\'ere for helpful discussion. \textbf{Funding:} This work is supported in part by the Air Force Office of Scientific Research under award number FA9550-19-1-0391. \textbf{Author contributions:} All authors designed the study. H.P. performed the analyses. H.P. and Y.A. produced the figures. H.P. and Y.A. led the writing of the manuscript. All authors contributed to the writing and approved the final manuscript. \textbf{Competing interests:} The authors declare that they have no competing interests. \textbf{Data and materials availability:} All code used in this study is available at: \url{https://github.com/haoopeng/periodicals}. The Microsoft Academic Graph data can be accessed at: \url{https://www.microsoft.com/en-us/research/project/microsoft-academic-graph/}. A public repository of our data is available at: \url{https://doi.org/10.6084/m9.figshare.13007650}.
\section*{Supplementary Materials}

Tables S1 to S3\\
Figures S1 to S13\\
Annotated map of journals in each discipline (Figures S14-S26)

\clearpage

\input{supp.tex}
\end{document}

%% file: supp.tex
\setcounter{figure}{0}
\setcounter{table}{0}
\renewcommand{\thefigure}{S\arabic{figure}}
\renewcommand{\thetable}{S\arabic{table}}
\renewcommand{\thesection}{S\arabic{section}}
\renewcommand{\theHtable}{Supplement.\thetable}
\renewcommand{\theHfigure}{Supplement.\thefigure}

\section{Supplementary Materials} 

\subsection{Supplementary Tables}



\begin{table}[ht] 
\begin{center}
\begin{small}
\begin{tabular}{|l|r|r|}
\hline
\textbf{Discipline Category} & \textbf{Num. of Journals} & \textbf{Percentage} \\ \hline
Biology&   1057 & 8.27 \\ \hline
Biotechnology &   238 & 1.86 \\ \hline
Brain Research &   741 & 5.80 \\ \hline
Chemical, Mechanical, \& Civil Engineering &  1023 & 8.00 \\ \hline
Chemistry &   644 & 5.04 \\ \hline
Earth Sciences &   490 & 3.83 \\ \hline
Electrical Engineering \& Computer Science &   779 & 6.10 \\ \hline
Health Professionals &  1387 & 10.85 \\ \hline
Humanities &   654 & 5.12 \\ \hline
Infectious Diseases &   660 & 5.16 \\ \hline
Math \& Physics &   738 & 5.77 \\ \hline
Medical Specialties &  1657 & 12.96 \\ \hline
Social Sciences &  2712 & 21.22 \\ \hline
Interdiscipline & 29 & 0.02 \\ \hline
\end{tabular}
\end{small}
\end{center}
\caption{The number of journals in 13 disciplines defined in the UCSD map of science. These 12,780 journals can be matched between the MAG data and the UCSD map data, and are covered in our embeddings. 29 journals belonging to multiple disciplines are labeled as ``Interdiscipline''. Throughout the paper, we abbreviate ``Chemical, Mechanical, \& Civil Engineering'' as ``Engineering'', and ``Electrical Engineering \& Computer Science'' as ``EE \& CS'' to save space.}
\label{tab:disc-match}
\end{table} 

\begin{table}[ht] 
\begin{center}
\begin{small}
\begin{tabular}{|c|c|c|c|c|}
\hline
$W$  & $D$   & $\Delta$Mean(\textbf{sub}) & $\Delta$Mean(\textbf{dis}) & Mean(\textbf{rand})     \\ \hline
2  & 50 &   0.302		   & 0.105                  & 0.233	    \\ \hline
2  & 100 & 0.349                  & 0.118                  & 0.253          \\ \hline
2  & 200 & 0.314                  & 0.103                  & 0.250          \\ \hline
2  & 300 & 0.299                  & 0.096                  & 0.243          \\ \hline
5  & 50 &   0.432		   & 0.179			& 0.073	     \\ \hline
5  & 100 & 0.457                  & 0.183                  & 0.086          \\ \hline
5  & 200 & 0.419                  & 0.165                  & 0.084          \\ \hline
5  & 300 & 0.399                  & 0.157                  & 0.082          \\ \hline
10 & 50  & 0.420		   & 0.172 		        & 0.069	     \\ \hline
\textbf{10} & \textbf{100} & \textbf{0.469}         & \textbf{0.192}         & 0.069 \\ \hline
10 & 200 & 0.428                  & 0.172                  & 0.067          \\ \hline
10 & 300 & 0.406                  & 0.161                  & 0.066          \\ \hline
\end{tabular}
\end{small}
\end{center}
\caption{\textbf{Hyperparameter tuning in the model training}. 
Each model is trained with the same 100 million periodical citation trails. 
The minimum frequency is set to 50 in all settings. 
$W$ is the context window size, $D$ is the number of embedding dimensions. 
Mean(\textbf{sub}), Mean(\textbf{dis}), and Mean(\textbf{rand}) are the mean cosine similarity of journal pairs in the same sub-discipline, journal pairs in the same discipline, and journal pairs in any discipline, respectively. Note that we randomly selected 100,000 journal pairs for each group. 
$\Delta$Mean(\textbf{sub}) $=$ Mean(\textbf{sub}) $-$ Mean(\textbf{rand}), $\Delta$Mean(\textbf{dis}) $=$ Mean(\textbf{dis}) $-$ Mean(\textbf{rand})).}
\label{tab:hyper}
\end{table} 

\begin{table}[ht]
\begin{center}
\begin{small}
\begin{tabular}{|l|r|r|}
\hline
\textbf{Discipline}           & \textbf{Num. of Participants} & \textbf{Num. of Selected Targets} \\ \hline
Social Sciences      & 50                   & 318                       \\ \hline
EE \& CS             & 39                   & 224                      \\ \hline
Engineering          & 20                   & 129                       \\ \hline
Math \& Physics      & 3                    & 21                       \\ \hline
Earth Sciences       & 2                    & 10                        \\ \hline
Health Professionals & 2                    & 9                        \\ \hline
Biology              & 1                    & 11                        \\ \hline
Brain Research       & 1                    & 5                        \\ \hline
Biotechnology        & 1                    & 4                        \\ \hline
\end{tabular}
\end{small}
\end{center}
\caption{The number of qualified participants and the total number of target journals selected by them across different disciplines.}
\label{tab:qualified}
\end{table}

\clearpage

\subsection{Supplementary Figures}

\begin{figure*}[ht!] 
\centering
\includegraphics[trim=0mm 0mm 0mm 0mm, width=0.6\columnwidth]{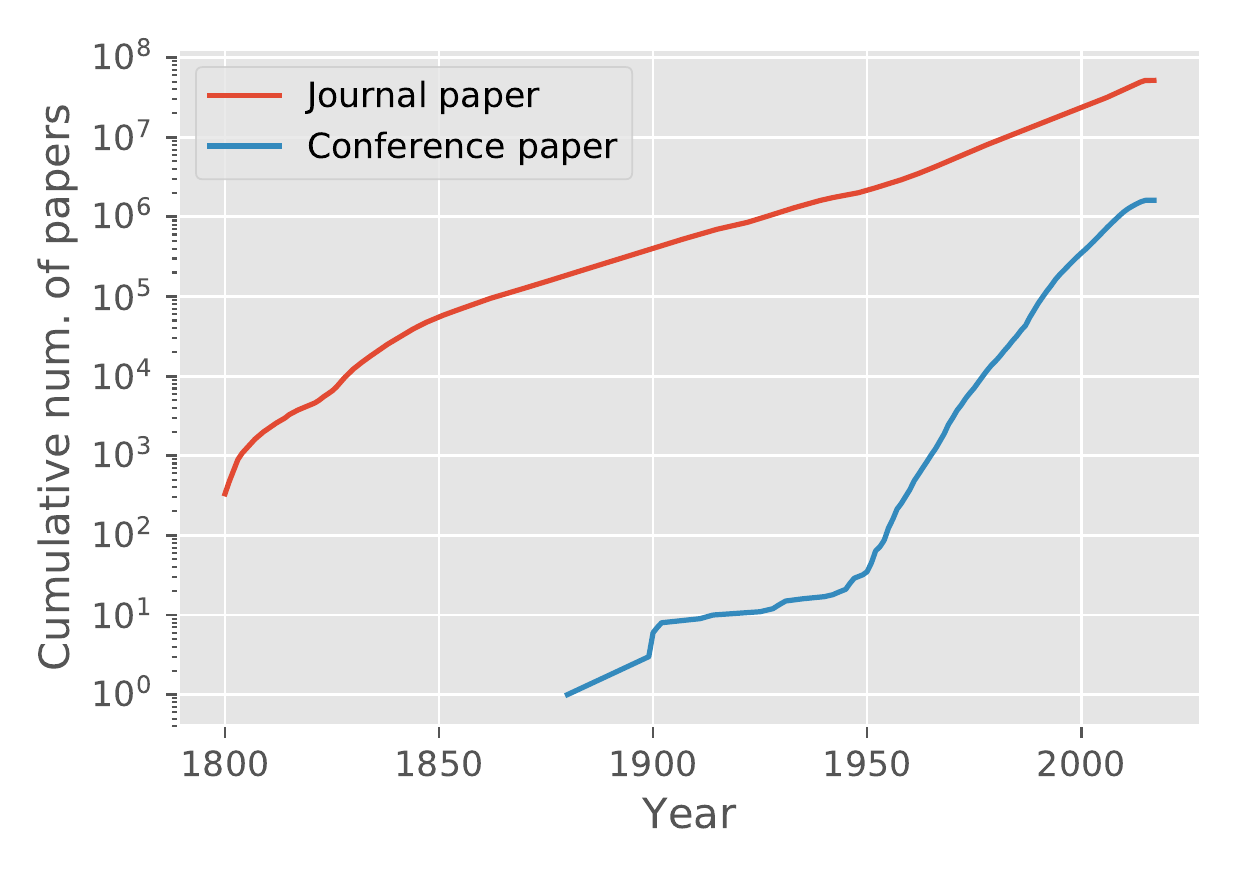}
\caption{The cumulative number of journal and conference papers from 1800 to 2016. A total number of 53 million papers are used in this study.}
\label{fig:count}
\end{figure*} 

\begin{figure*}[ht!] 
\centering
\includegraphics[trim=0mm 0mm 0mm 0mm, width=0.6\columnwidth]{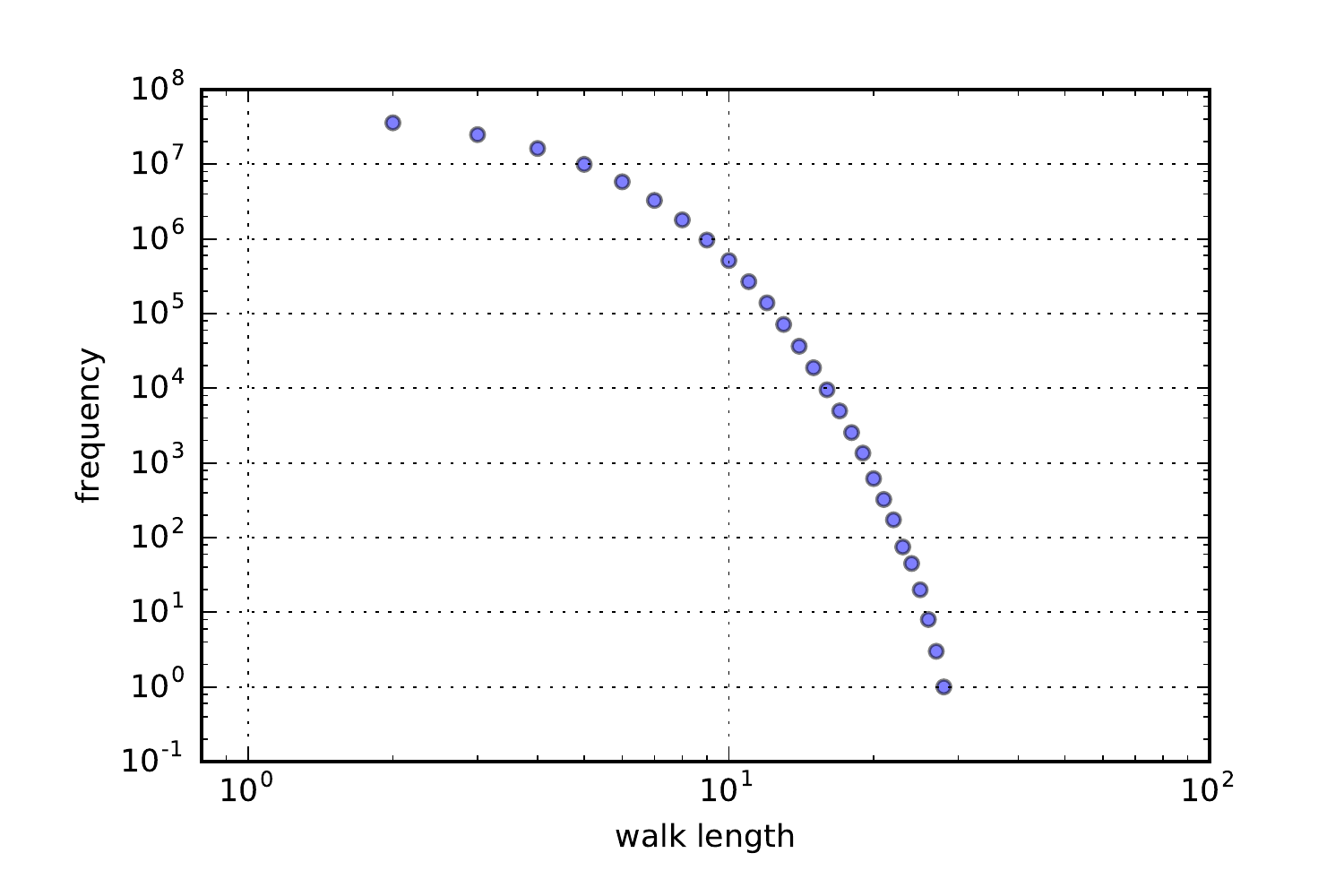}
\caption{The length distribution of 100 million periodical trails. Note that length-one trails were discarded during the random walk process.}
\label{fig:walk-length}
\end{figure*} 



\begin{figure*}[ht!] 
\centering
\includegraphics[trim=0mm 0mm 0mm 0mm, width=0.5\linewidth]{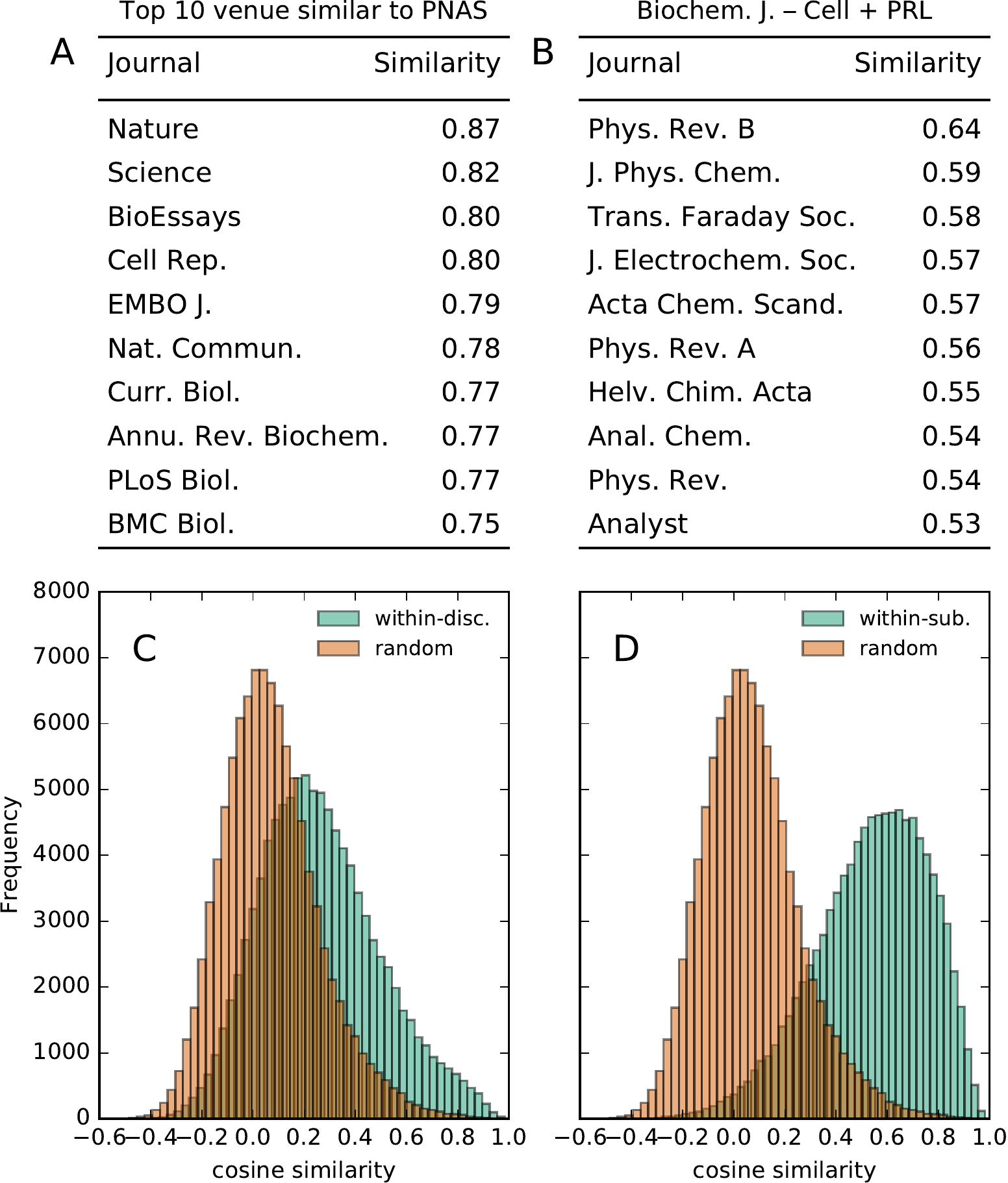}
\caption{\textbf{Periodical recommendations.}
(A) The 10 most similar periodicals to \emph{PNAS}, a multi-disciplinary yet biomedical-dominated journal, based on the cosine similarities between periodical embeddings. Other multi-disciplinary journals, such as \emph{Nature}, \emph{Science}, \emph{Nature Communications}, and some biological journals are among the top list. 
(B) The 10 most similar periodicals to the vector analogy: $\mathbf{v}(\emph{Biochemical Journal}) - \mathbf{v}(\emph{Cell}) + \mathbf{v}(\emph{Physical Review Letters})$. 
(C) The histogram of cosine similarities of 100,000 randomly selected journal pairs that are in the same discipline (``within-disc.'') or in any discipline (``random''). 
(D) As in (C), but for journal pairs in the same sub-discipline (``within-sub.'').}
\label{fig:simi}
\end{figure*} 

\begin{figure*}[ht] 
\centering
\includegraphics[trim=0mm 0mm 0mm 0mm, width=0.8\linewidth]{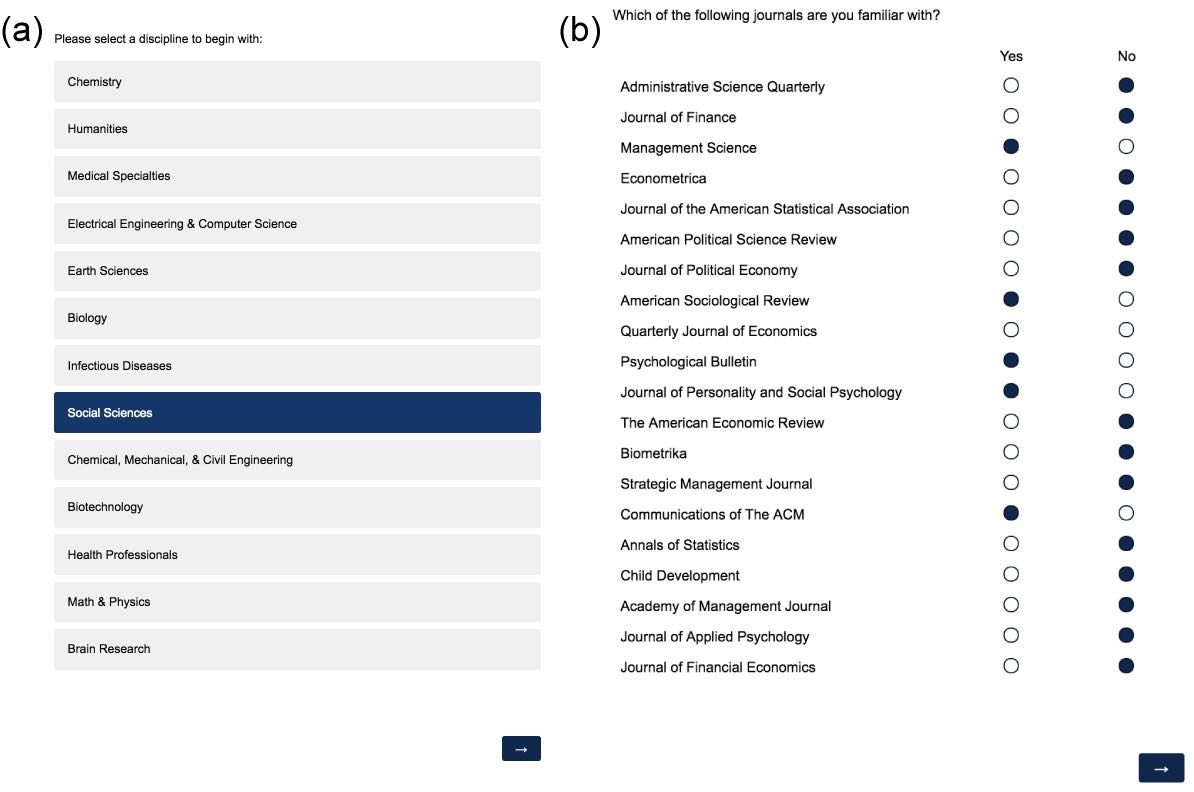}
\caption{\textbf{The interface of the journal recommendation survey.} \textbf{a}, The survey interface where participants were first asked to choose a discipline to begin the task. \textbf{b}, Participants were asked about their familiarity with the 20 target journals in ``Social Sciences''. The survey continues only if at least 3 target journals were selected.}
\label{fig:survey-one}
\end{figure*} 

\begin{figure*}[ht] 
\centering
\includegraphics[trim=0mm 0mm 0mm 0mm, width=0.8\linewidth]{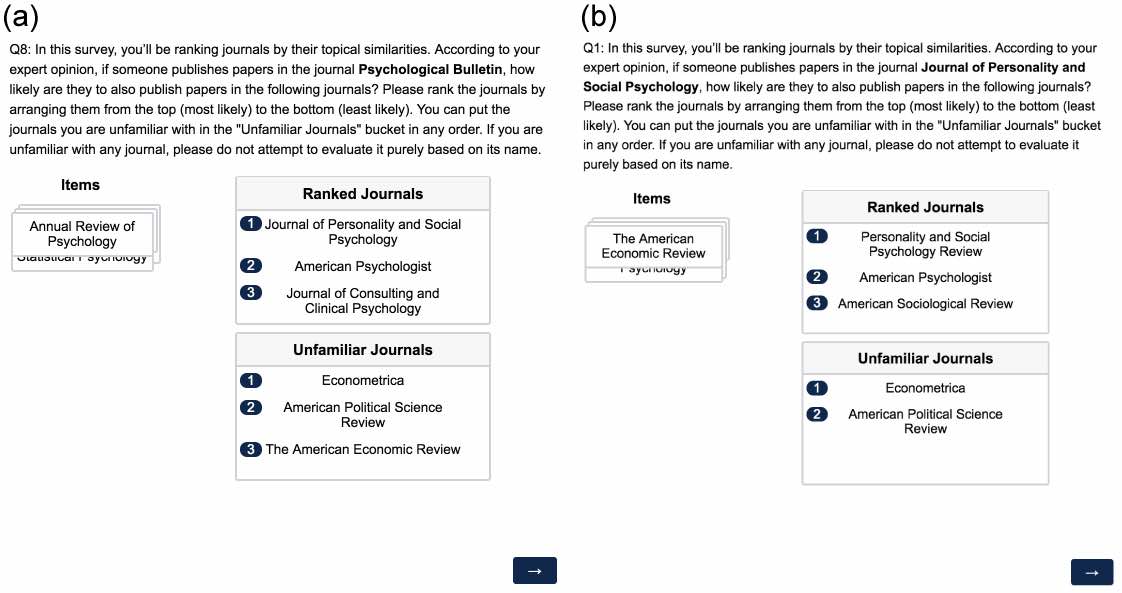}
\caption{\textbf{Screenshots of the rank task interface for two exemplar target journals.} \textbf{a}, \emph{Psychological Bulletin}. \textbf{b}, \emph{Journal of Personality and Social Psychology}. The candidate journals on the left side are randomly stacked on top of each other. Participants can place unfamiliar candidates in the ``Unfamiliar Journals'' bucket in any order.}
\label{fig:survey-two}
\end{figure*} 

\begin{figure*}[ht] 
\centering
\includegraphics[trim=0mm 0mm 0mm 0mm, width=0.7\linewidth]{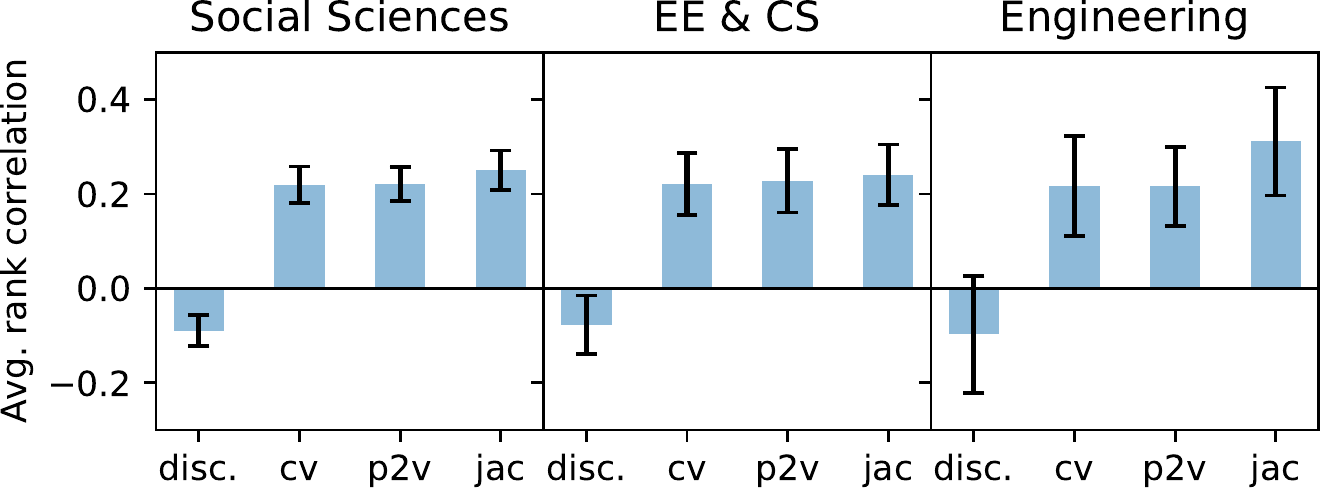}
\caption{\textbf{The average Kendall's rank correlation coefficient between experts and four models in three disciplines.} Target journals with an average expert agreement above 0.2 are used in the evaluation. The four labels---\textit{disc.}, \textit{cv}, \textit{jac}, and \textit{p2v}---represent the first baseline method, the citation-based sparse vector-space model, the Jaccard similarity matrix, and our periodical embeddings.}
\label{fig:survey-j-three-disc}
\end{figure*} 

\begin{figure*}[ht!] 
\centering
\includegraphics[trim=0mm 0mm 0mm 0mm, width=\linewidth]{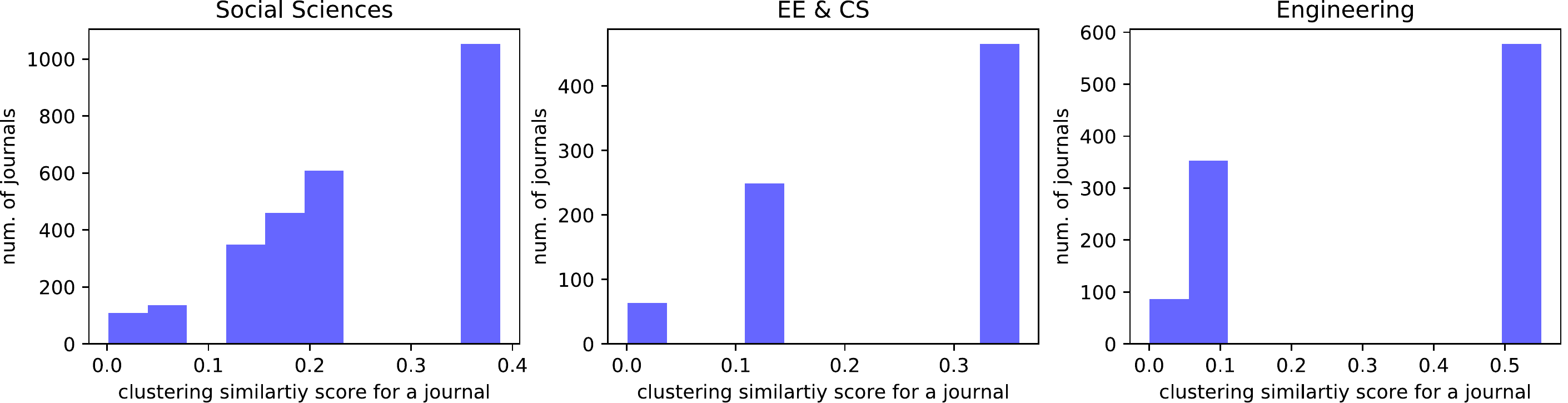}
\caption{Histograms of the similarity (agreement) scores between two clusterings (UCSD journal categorizations vs. a clustering based on our periodical embeddings) for journals in three disciplines, which display a multimodal distribution. The discreteness comes from the fact that there are only 13 clusters and we are comparing two clusterings.}
\label{fig:agree-hist}
\end{figure*} 


\begin{figure*}[ht!] 
\centering
\includegraphics[trim=0mm 0mm 0mm 0mm, width=0.35\linewidth]{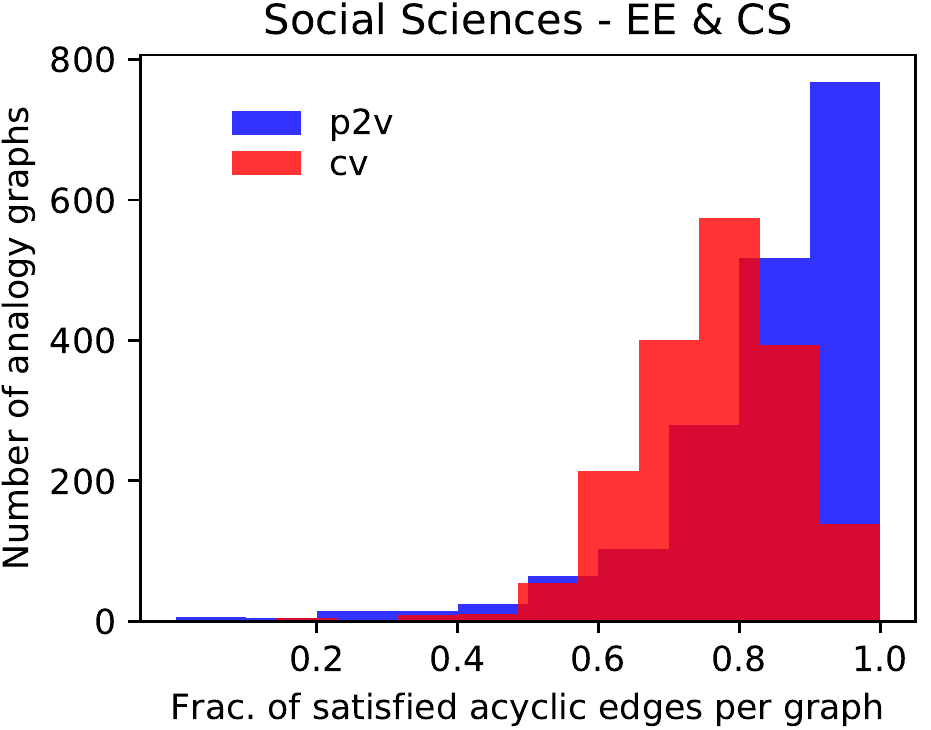}
\caption{The distribution of the fraction of acyclic edges that satisfy the author overlap criterion across 1,800 analogy graphs in (``Social Sciences'', ``EE \& CS''). For a periodical analogy ``$A : B \sim C : D$'', we verify whether $\frac{O(C, A)}{O(C, B)} > \frac{O(D, A)}{O(D, B)}$. Here $O(P_1, P_2)$ indicates the number of common authors who have ever published a paper in both periodicals. Our embedding method (\textit{p2v}) is compared against the citation-based sparse encoding model (\textit{cv}) for generating analogy graphs. The mean value for \textit{p2v} is significantly higher than that for \textit{cv} (0.84 vs 0.76).}
\label{fig:analogy-a}
\end{figure*} 

\begin{figure*}[ht!] 
\centering
\includegraphics[trim=0mm 0mm 0mm 0mm, width=0.7\linewidth]{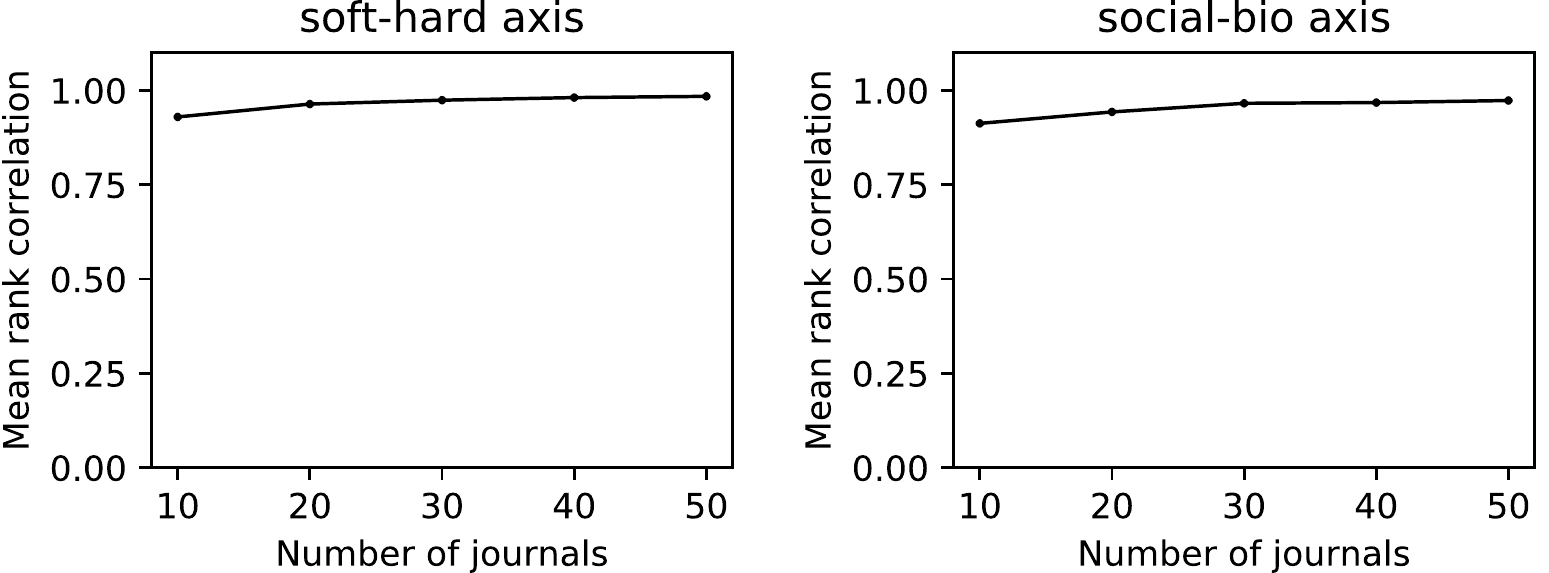}
\caption{The average Spearman's rank correlation between the ordering of 20,835 periodicals (covered in our embedding model) and their ordering based on an axis built with a subset of randomly selected journals in the two broad disciplines. The $x$-axis represents the number of journals in the subset. Error bars indicate 95\% confidence intervals.}
\label{fig:axes-stability}
\end{figure*} 

\begin{figure*}[ht!]
\centering
\includegraphics[trim=0mm 0mm 0mm 0mm, width=\linewidth]{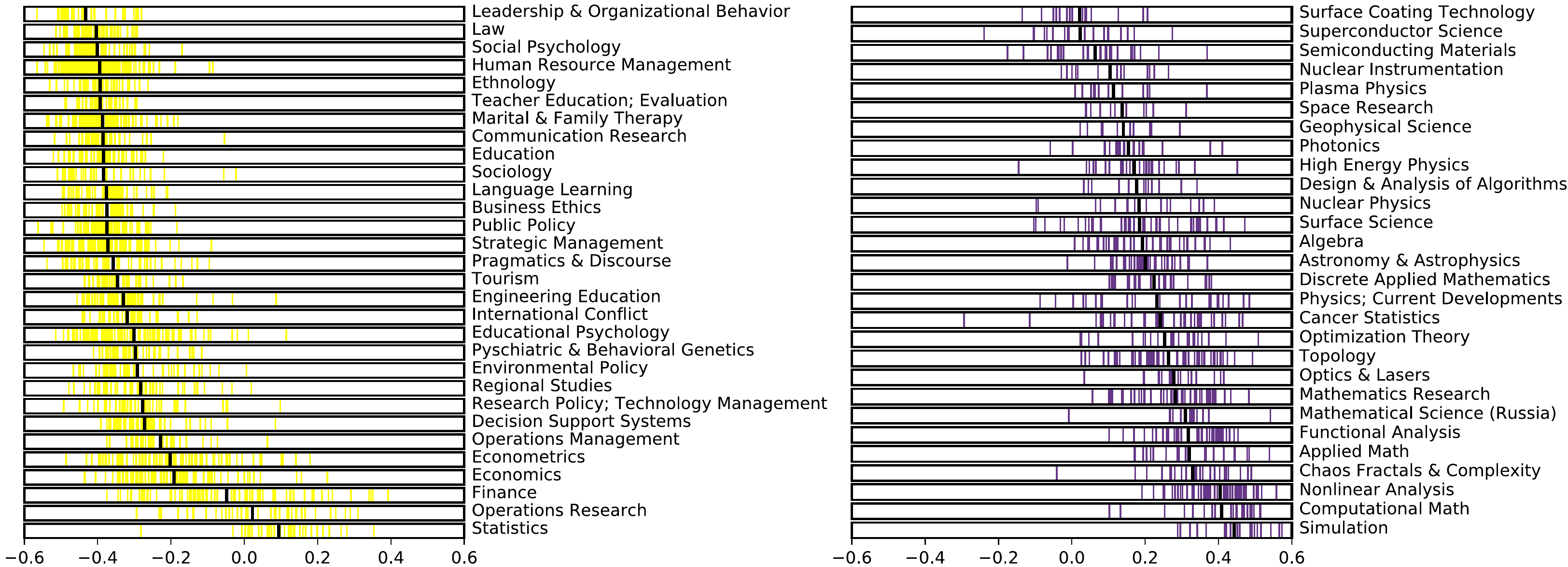}
\caption{\textbf{The organization of sub-disciplines on the ``soft-hard'' sciences axis.} \textbf{Left,} The spectrum of sub-disciplines in ``Social Sciences'' on this axis. Each journal is represented by a vertical line inside the box. Sub-disciplines are ordered by their mean values (the black vertical line). There is a relatively clear separation of subfields on this axis.
\textbf{Right,} The spectrum of sub-disciplines in ``Math \& Physics'' on the ``soft-hard'' sciences axis, which does not exhibit a separation as clear as that in ``Social Sciences''. Note that, to save space, only the top 30 subfields (based on their number of journals) are shown in each discipline.}
\label{fig:soft-hard}
\end{figure*} 

\begin{figure*}[ht!]
\centering
\includegraphics[trim=0mm 0mm 0mm 0mm, width=\linewidth]{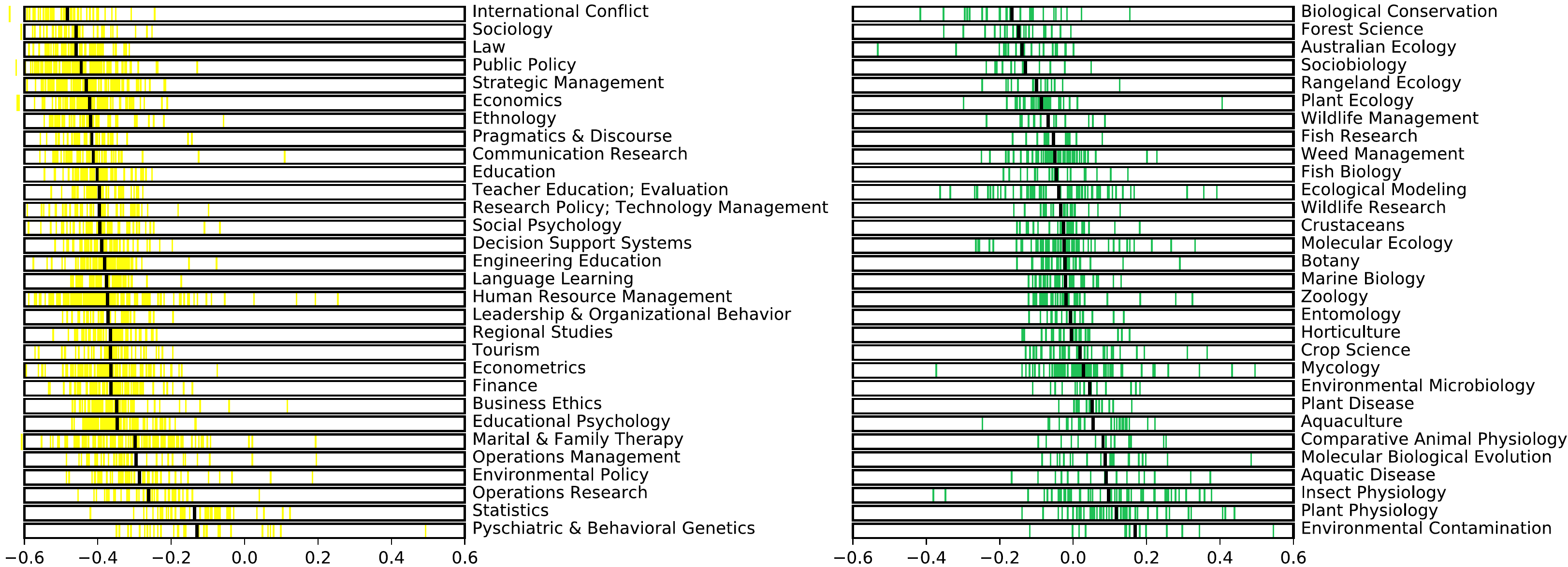}
\caption{\textbf{The organization of sub-disciplines on the ``social-bio'' science axis.} \textbf{Left}, The spectrum of sub-disciplines in ``Social Sciences'' on this axis. Each journal is represented by a vertical line inside the box. Sub-disciplines are ordered by their mean values (the black vertical line). There is a relatively clear separation of subfields on this axis.
\textbf{Right}, The spectrum of sub-disciplines in ``Biology'' on the ``social-bio'' science axis, which does not exhibit a separation as clear as that in ``Social Sciences''. Note that only the top 30 subfields (based on their number of journals) are shown in each discipline.}
\label{fig:social-bio}
\end{figure*} 



\begin{figure*}[ht!] 
\centering
\includegraphics[trim=0mm 0mm 0mm 0mm, width=\linewidth]{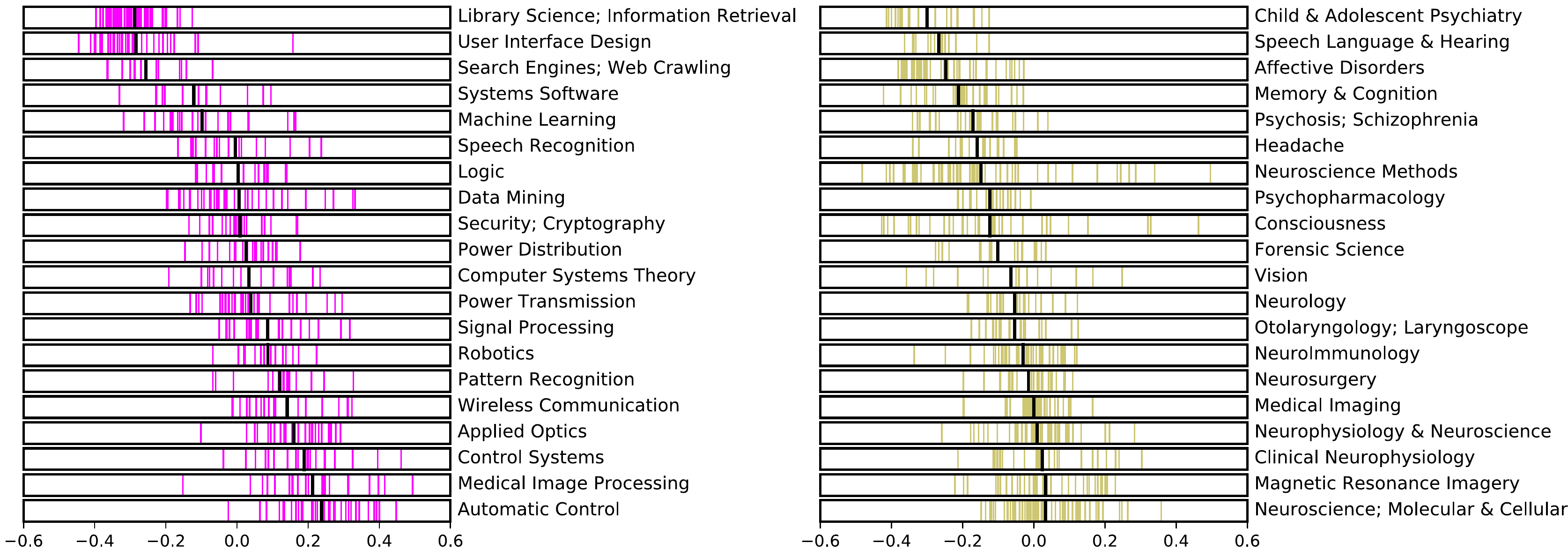}
\caption{\textbf{The ordering of sub-disciplines on the ``soft-hard'' sciences axis.} \textbf{Left,} The spectrum of journals in sub-disciplines of ``\textbf{EE \& CS}'' on the ``soft-hard'' sciences axis. Each journal is represented by a vertical line inside the box. The color represents the discipline category in the UCSD map of science. We focus on the top 20 sub-disciplines based on the number of journals, and ordered each category by their mean projection values (the black vertical line). Research domains such as ``Library Science'', ``Information Retrieval'', ``User Interface Design'', ``Machine Learning'', and ``Data Mining'' are ``softer'' than domains such as ``Signal Processing'', ``Robotics'', ``Wireless Communication'', and ``Controls Systems''. \textbf{Right,} The spectrum of journals in ``\textbf{Brain Research}'' on the ``soft-hard'' sciences axis. Sub-disciplines such as ``Neurology'', ``Medical Imaging'', and ``Magnetic Resonance Imagery'' are ``harder'' than ``Psychiatry'', ``Speech'', ``Hearing'', ``Headache'', and ``Consciousness''.}
\label{fig:spectrum-three}
\end{figure*} 


\begin{figure*}[ht!] 
\centering
\includegraphics[trim=0mm 0mm 0mm 0mm, width=0.5\linewidth]{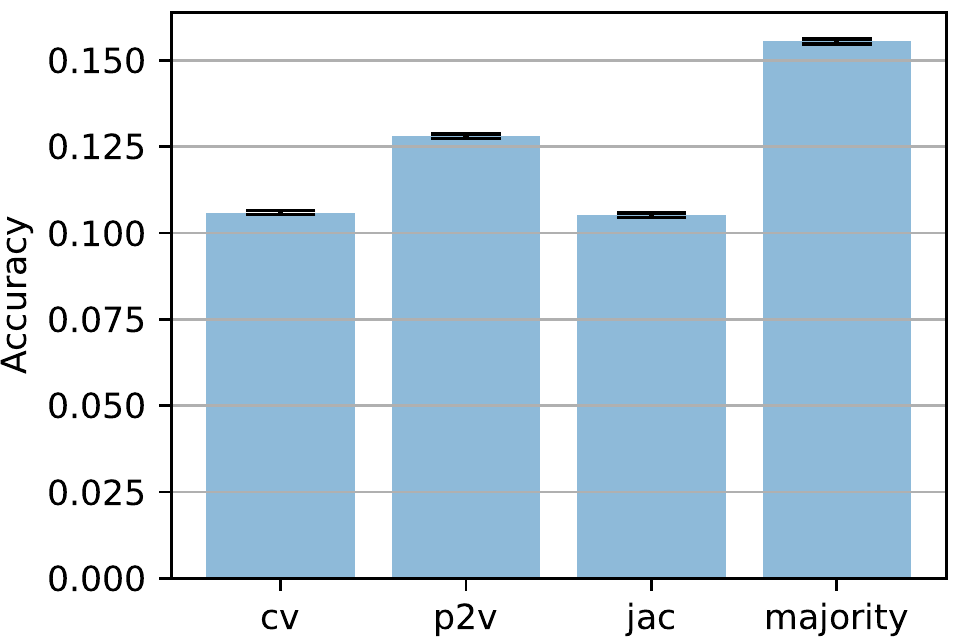}
\caption{\textbf{The accuracy of four different methods in predicting papers' publication venue based on their cited periodicals.} We use a random sample of 10,000 papers published after year 2000 (as younger papers are likely to have more complete reference information in the MAG dataset). The experiment is repeated 100 times to estimate the error. Our dense periodical embedding model (\textit{p2v}) is compared with the citation vector model (\textit{cv}), the Jaccard similarity matrix (\textit{jac}), and another baseline method (\textit{majority}) that predicts a paper's publication venue to be the most frequently cited periodical in its reference list. The three vector-space models predict the periodical that is the closest to the average vector of cited periodicals based on cosine similarity.}
\label{fig:pred-pub-venue}
\end{figure*} 


\clearpage

\subsection{Annotated map of journals in each discipline} \label{annotation}

\begin{figure*}[ht!] 
\centering
\includegraphics[trim=0mm 0mm 0mm 0mm, width=\linewidth]{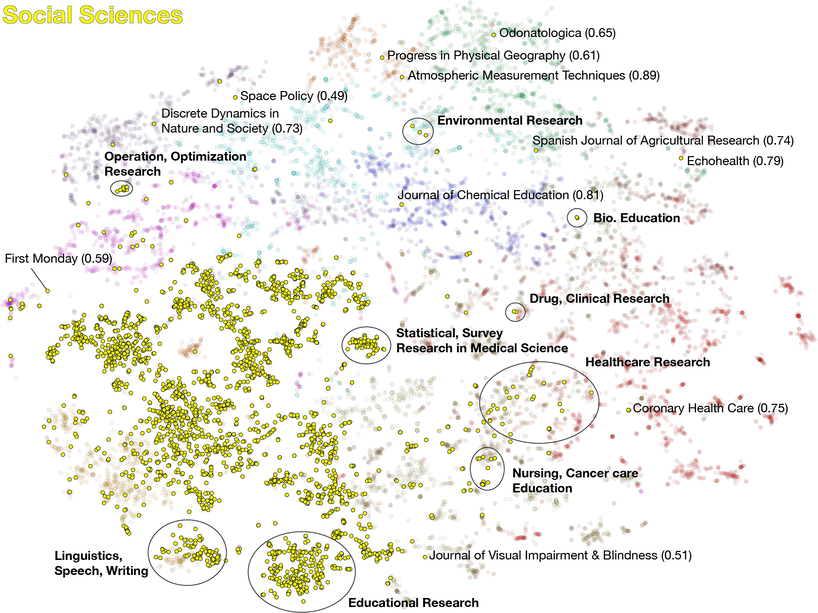}
\caption{The realm of ``Social Sciences'' journals in the embedding space.}
\label{fig:map-social}
\end{figure*}

\begin{figure*}[ht!] 
\centering
\includegraphics[trim=0mm 0mm 0mm 0mm, width=\linewidth]{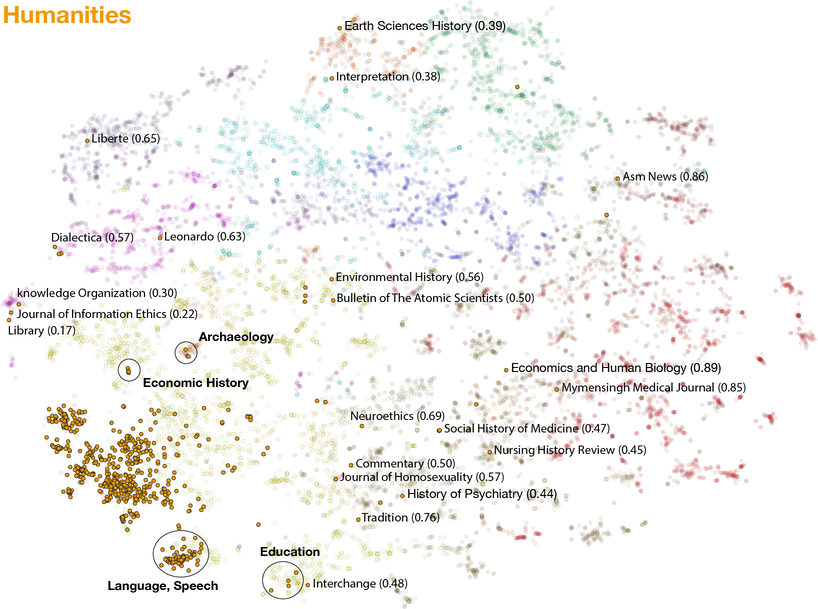}
\caption{The region of ``Humanities'' journals in the embedding space.}
\label{fig:map-humanities}
\end{figure*}

\begin{figure*}[ht!] 
\centering
\includegraphics[trim=0mm 0mm 0mm 0mm, width=\linewidth]{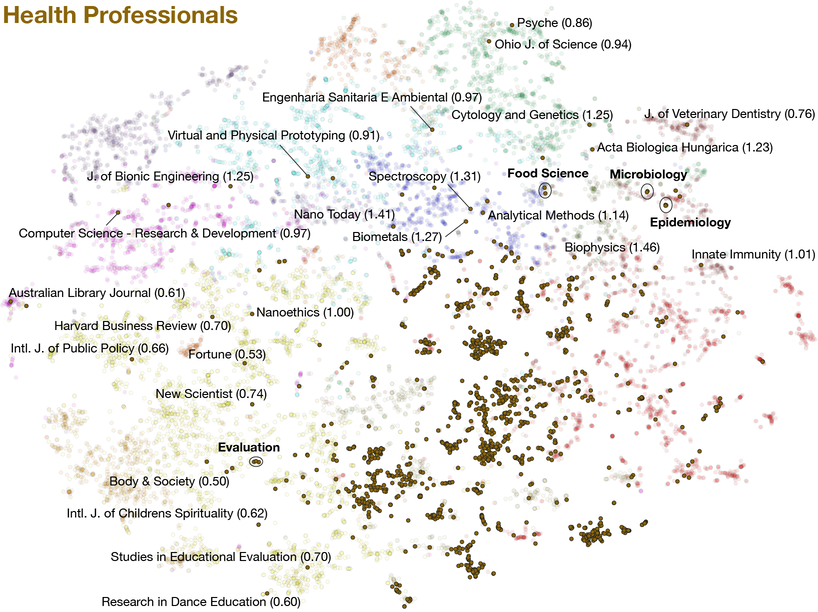}
\caption{The colony of ``Health Professionals'' journals in the embedding space.}
\label{fig:map-health}
\end{figure*}

\begin{figure*}[ht!] 
\centering
\includegraphics[trim=0mm 0mm 0mm 0mm, width=\linewidth]{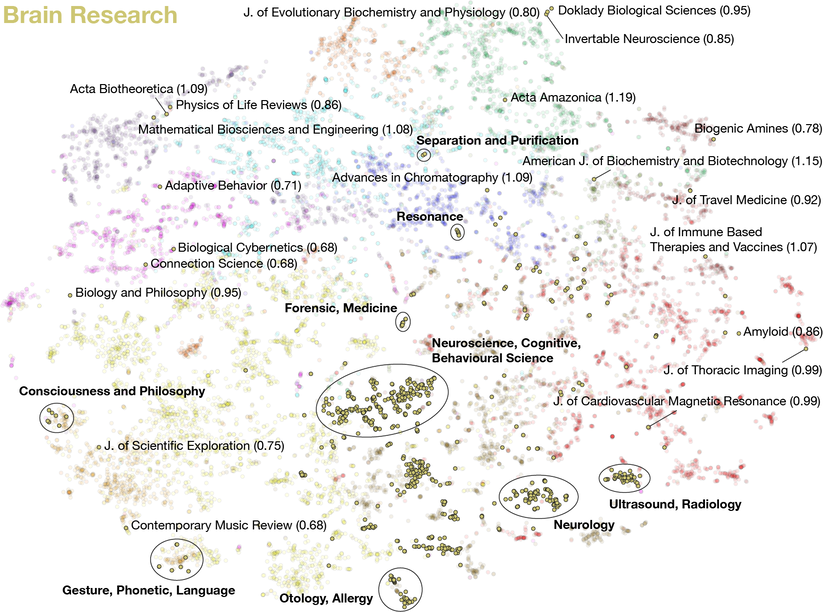}
\caption{The realm of ``Brain Research'' journals in the embedding space.}
\label{fig:map-brain}
\end{figure*}

\begin{figure*}[ht!] 
\centering
\includegraphics[trim=0mm 0mm 0mm 0mm, width=\linewidth]{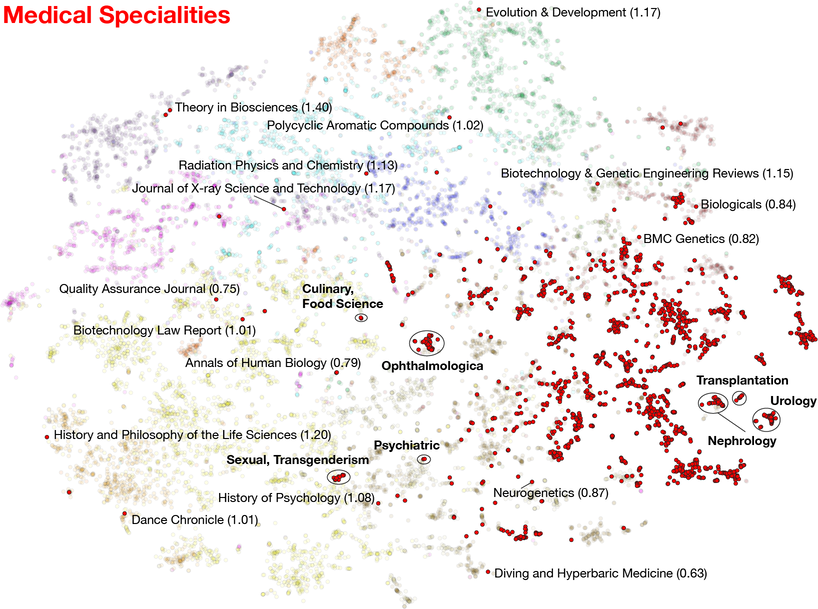}
\caption{The region of ``Medical Specialties'' journals in the embedding space.}
\label{fig:map-medical}
\end{figure*}

\begin{figure*}[ht!] 
\centering
\includegraphics[trim=0mm 0mm 0mm 0mm, width=\linewidth]{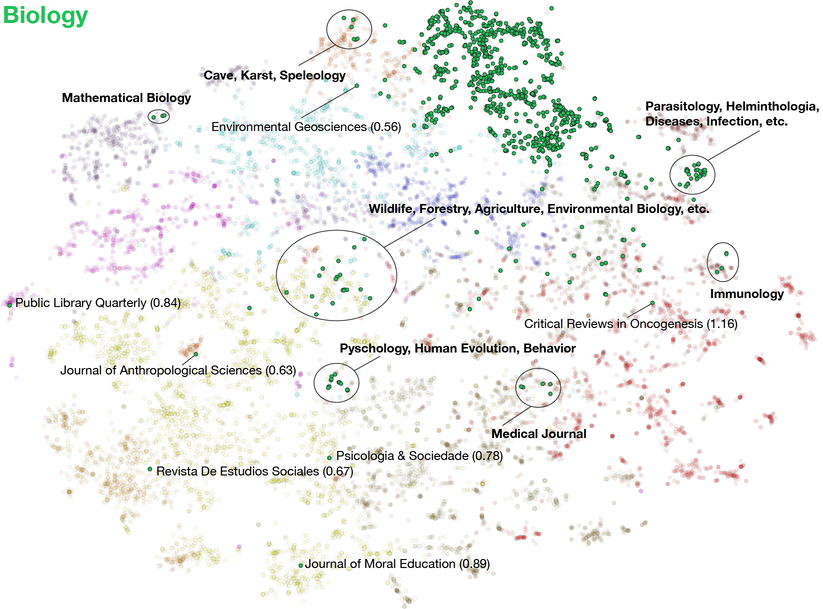}
\caption{The colony of ``Biology'' journals in the embedding space.}
\label{fig:map-bio}
\end{figure*}

\begin{figure*}[ht!] 
\centering
\includegraphics[trim=0mm 0mm 0mm 0mm, width=\linewidth]{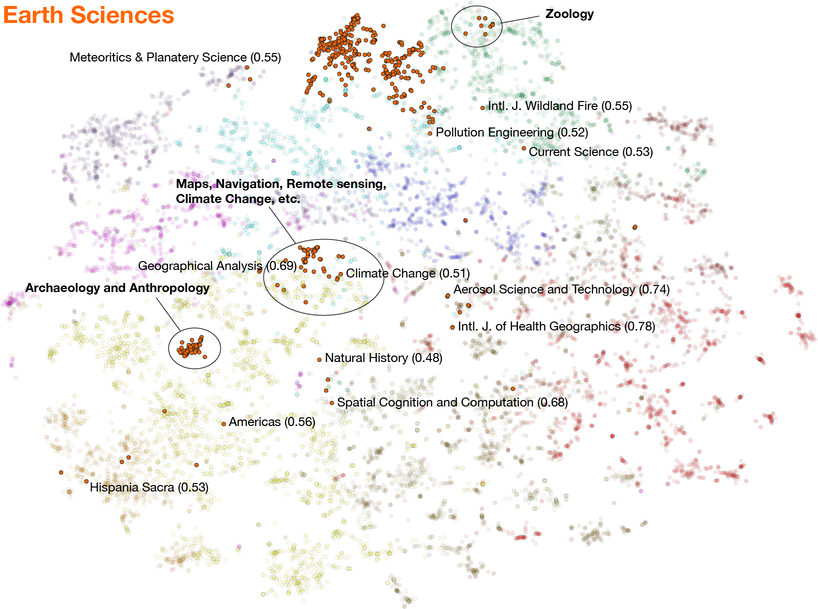}
\caption{The realm of ``Earth Sciences'' journals in the embedding space.}
\label{fig:map-earth}
\end{figure*}

\begin{figure*}[ht!] 
\centering
\includegraphics[trim=0mm 0mm 0mm 0mm, width=\linewidth]{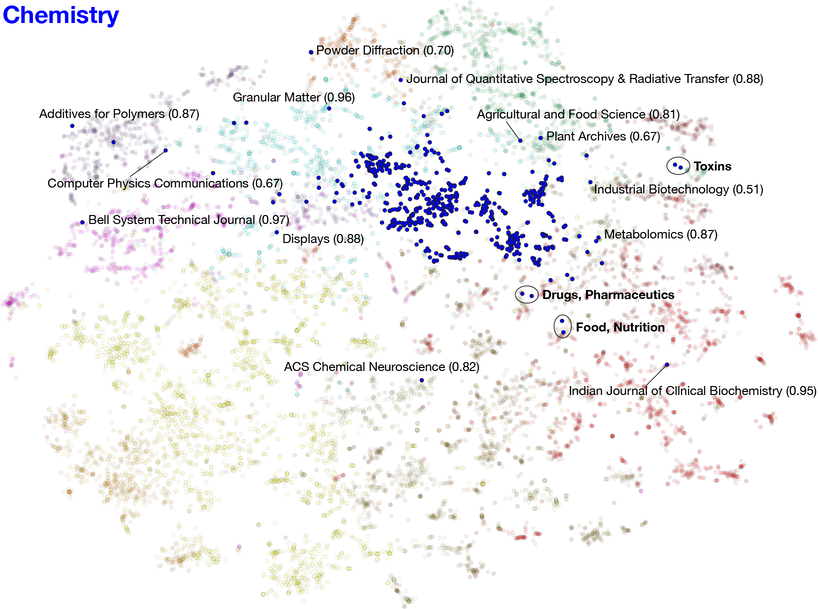}
\caption{The region of ``Chemistry'' journals in the embedding space.}
\label{fig:map-chem}
\end{figure*}

\begin{figure*}[ht!] 
\centering
\includegraphics[trim=0mm 0mm 0mm 0mm, width=\linewidth]{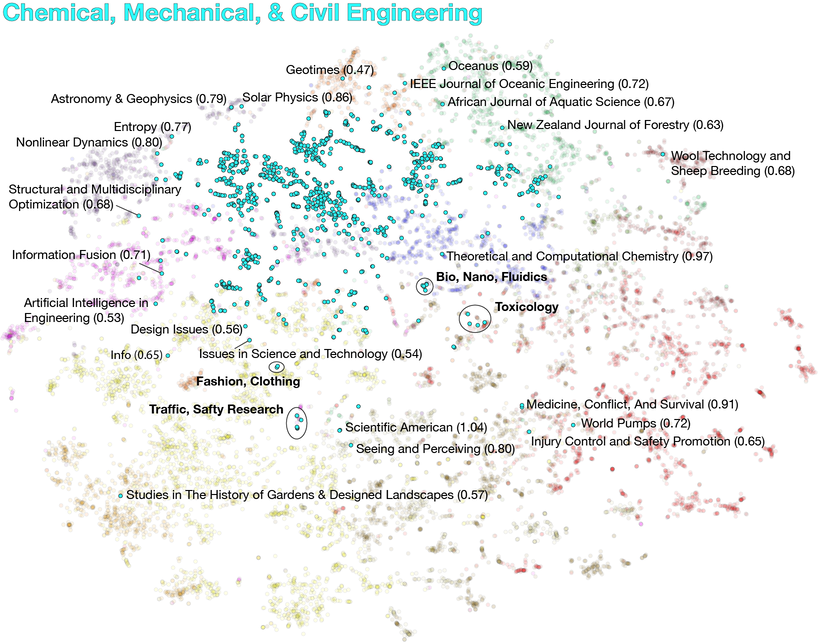}
\caption{The colony of ``Chemical, Mechanical, \& Civil Engineering'' journals.}
\label{fig:map-eng}
\end{figure*}

\begin{figure*}[ht!] 
\centering
\includegraphics[trim=0mm 0mm 0mm 0mm, width=\linewidth]{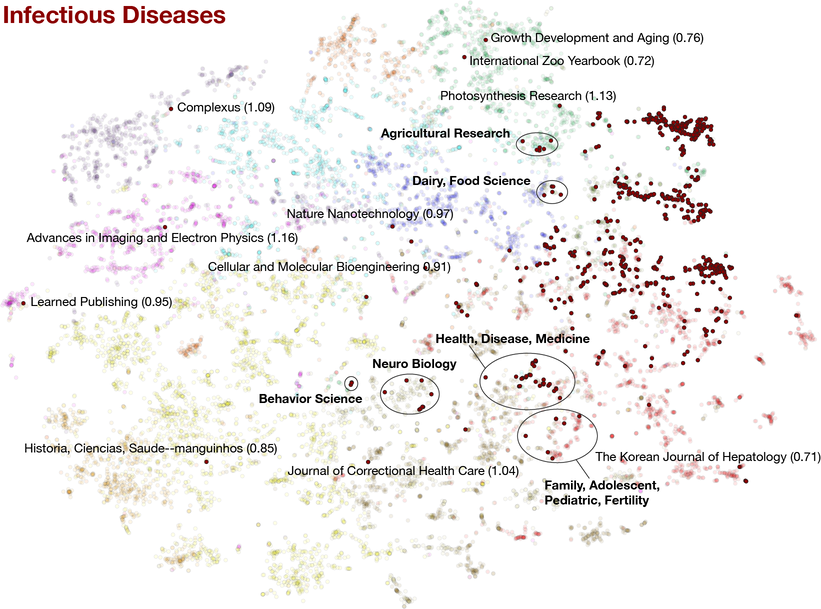}
\caption{The realm of ``Infectious Diseases'' journals in the embedding space.}
\label{fig:map-diseases}
\end{figure*}

\begin{figure*}[ht!] 
\centering
\includegraphics[trim=0mm 0mm 0mm 0mm, width=\linewidth]{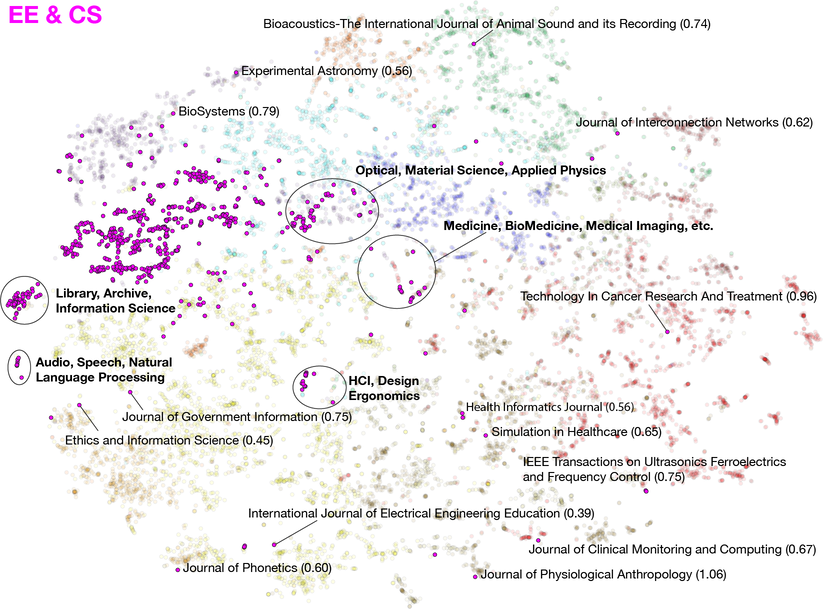}
\caption{The region of ``Electrical Engineering \& Computer Science'' journals.}
\label{fig:map-eecs}
\end{figure*}

\begin{figure*}[ht!] 
\centering
\includegraphics[trim=0mm 0mm 0mm 0mm, width=\linewidth]{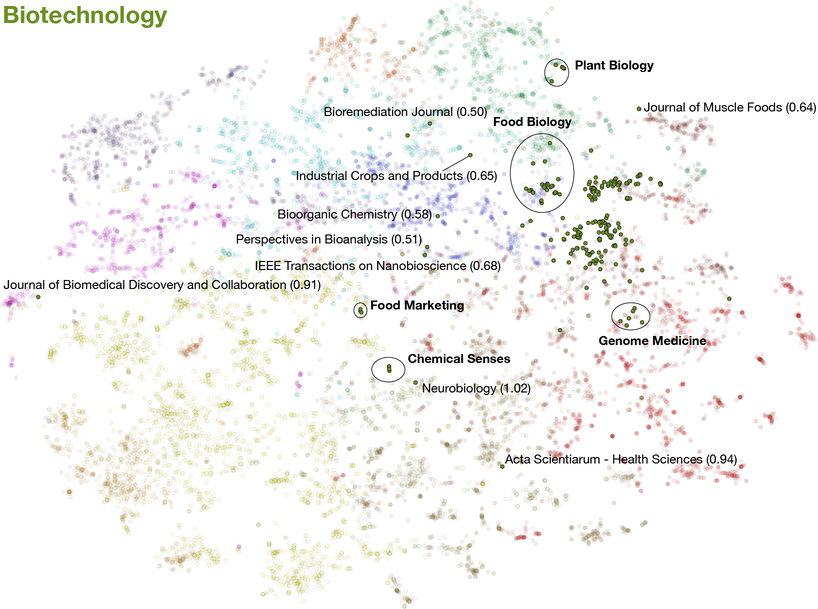}
\caption{The colony of ``Biotechnology'' journals in the embedding space.}
\label{fig:map-biotech}
\end{figure*}

\begin{figure*}[ht!] 
\centering
\includegraphics[trim=0mm 0mm 0mm 0mm, width=\linewidth]{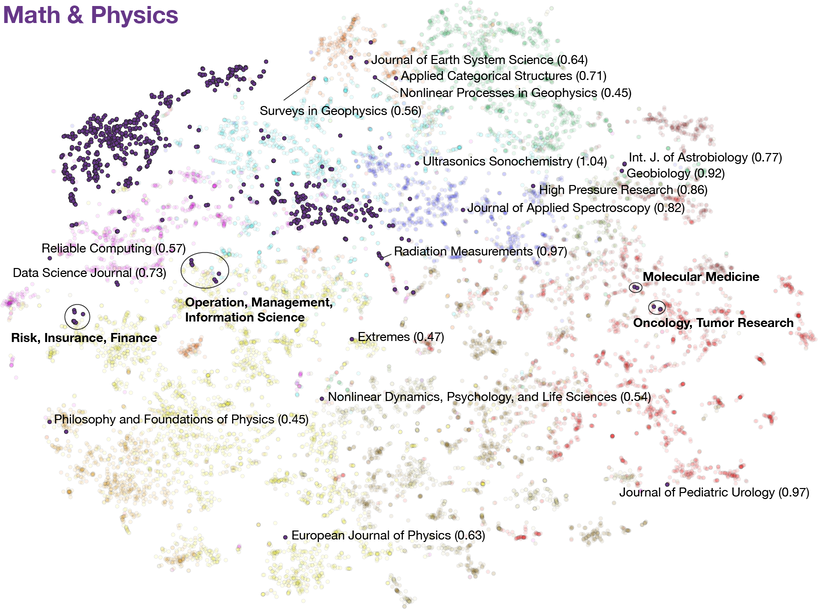}
\caption{The territory of ``Math \& Physics'' journals in the embedding space.}
\label{fig:map-mathphys}
\end{figure*}

The 2-\emph{d} projection of the embeddings of 12,780 journals provides an overview of the organizational structure of major academic disciplines. 
Here we further investigate the interdisciplinary nature of many academic journals in each discipline and show that many of them cannot be properly categorized into a unique discipline by an existing journal classification system---the UCSD map of science.
Figs.~\ref{fig:map-social}--\ref{fig:map-mathphys} highlight all journals in a given discipline with journals in all other disciplines blurred in the background. The discipline category comes from the UCSD catalog.
In the map of each discipline, we annotated some exemplar interdisciplinary journals and micro-clusters that are located near disciplinary boundaries or are far away from their main discipline clusters. 
When annotating individual journals, we also provide its cosine distance to the centroid of all journal vectors in that discipline. Journals that are far away from its discipline center are likely to be misclassified in the UCSD map of science.

